\documentclass{aa}  
\usepackage{graphicx}
\usepackage{amsmath,amssymb}
\usepackage{txfonts}
\begin{document}

\def\Arrow{\mathop{\longrightarrow}\limits}
\def\Harpoons{\mathop{\rightleftharpoons}\limits}

   \title{Prestellar and protostellar cores in Ori B9
   \thanks{This publication is based on data acquired with the IRAM 30 m
telescope and the Atacama Pathfinder Experiment (APEX). IRAM is supported by
INSU/CNRS (France), MPG (Germany), and IGN (Spain). APEX is a collaboration
between the Max-Planck-Institut f\"ur Radioastronomie, the European Southern
Observatory, and the Onsala Space Observatory.}}

   \author{O. Miettinen\inst{1}, J. Harju\inst{1}, L. K. Haikala\inst{1}, 
           J. Kainulainen\inst{1,2}, \and L. E. B. Johansson\inst{3}}

   \offprints{O. Miettinen}

   \institute{Observatory, P.O. Box 14, 00014 University of Helsinki, Finland\\
              \email{oskari.miettinen@helsinki.fi} \and TKK/Mets\"ahovi Radio Observatory, Mets\"ahovintie 114, FIN-02540 Kylm\"al\"a, Finland \and Onsala Space Observatory, SE-439 92 Onsala, Sweden}

\date{Received ; accepted}

\authorrunning{Miettinen et al.}
\titlerunning{N$_2$H$^+$, N$_2$D$^+$ and dust emission in Ori B9}

  \abstract
   {Dense molecular cores are studied in order to gain
insight into the processes causing clouds to fragment and form stars.
In this study we concentrate on a region which is assumed to 
represent an early stage of clustered star formation in a giant molecular
cloud.}
   {The aims of this study are to determine the properties and
spatial distribution of dense cores in the relatively quiescent Ori B9
cloud, and to estimate their ages and dynamical timescales.} 
   {The cloud was mapped in the 870 $\mu$m continuum with
APEX/LABOCA, and selected positions were observed in the lines of N$_2$H$^+$
and N$_2$D$^+$ using IRAM-30m. These were used together with our previous
H$_2$D$^+$ observations to derive the degree of deuteration and some other
chemical characteristics. Moreover, archival far-infrared Spitzer/MIPS maps 
were combined with the LABOCA map to distinguish between prestellar and
protostellar cores, and to estimate the evolutionary stages of protostars.}
   {Twelve dense cores were detected at 870 $\mu$m continuum in
the Ori B9 cloud. The submm cores constitute $\sim4\%$ of the total
mass of the Ori B9 region.
There is an equal number of prestellar and protostellar cores. Two of the
submm sources, which we call SMM 3 and SMM 4, are previously unknown Class 0
candidates. There is a high likelyhood that the core masses and mutual
separations represent the same distributions as observed in other parts of
Orion. We found a moderate degree of deuteration in N$_2$H$^+$ ($0.03-0.04$).
There is, furthermore, evidence for N$_2$H$^+$ depletion in the core SMM 4.
These features suggest the cores have reached chemical maturity.
We derive a relatively high degree of ionization
($\sim 10^{-7}$) in the clump associated with IRAS 05405-0117. The ambipolar
diffusion timescales for two of the cores are $\sim70-100$ times longer than
the free-fall time.} 
   {The distribution and masses of dense cores in Ori B9 are
similar to those observed in more active regions in Orion, where the
statistical core properties have been explained by turbulent
fragmentation. The 50/50 proportions of prestellar and protostellar
cores suggest that duration of the prestellar phase is comparable
to the free-fall time. This timescale can be questioned, however, on
the basis of chemical data on the IRAS 05405-0117 region. A possible
explanation is that this survey samples only the densest, i.e., 
dynamically most advanced cores.}


    \maketitle
%

\section{Introduction}

Most stars form in clusters and smaller groups in the densest parts of giant
molecular clouds (GMCs). The fragmentation of molecular
clouds results in dense filaments which contain still denser cores.
These cold star-forming cores are best detected using far-infrared (FIR) and
submillimetre (submm) dust continuum. By studying their physical and
chemical characteristics one hopes to understand conditions leading to
protostellar collapse and the timescale related to this process. Furthermore,
the distribution and spacing of dense cores place constraints on the
fragmentation mechanisms (e.g., turbulent fragmentation and ambipolar
diffusion) and the possible interaction between newly born stars and their
surroundings (e.g., \cite{megeath2008}).
The estimates of the core masses resulting from dust continuum data can also be
used to examine the possible connection between the mass distribution of dense
cores and the stellar initial mass function (IMF), a question of great current
interest (e.g., \cite{simpson2008}; \cite{swift2008}; \cite{goodwin2008}; 
Rathborne et al. 2009, submitted).

Parameters affecting the cloud dynamics, such as the
degree of ionization and the abundances of various positive ions are
chemically related to deuterium fractionation and depletion of heavy
species (e.g., CO). Besides being important for the core
dynamics through the magnetic support and molecular line cooling,
these parameters depend on the core
history and characterise its present evolutionary stage. 
For example, substantial CO depletion and deuterium
enrichment are supposed to be a characteristic
of prestellar cores in the pivotal stage before collapse 
(\cite{caselli1999}; \cite{bacmann2002}; \cite{lee2003}).
Observations (\cite{tafalla2002}, 2004, 2006) and some theoretical models
(\cite{bergin1997}; \cite{aikawa2005}) suggest, however, that N-containing
species such as the chemically closely related nitrogen
species N$_2$H$^+$ and NH$_3$ (and their deuterated isotopologues),
remain in the gas phase at densities for which CO and other C-containing
molecules are already depleted (e.g., \cite{flower2005}; 2006b and references
therein).
They are therefore considered as useful spectroscopic tracers of prestellar
cores and the envelopes of protostellar cores.
There is, however, some evidence that N$_2$H$^+$ finally
freezes out at densities $n({\rm H_2})=$ several $\times10^5$ to $\gtrsim10^6$ 
cm$^{-3}$ (e.g., \cite{bergin2002}; \cite{pagani2005}, 2007). Contrary 
to this, NH$_3$ abundance appears to \textit{increase} toward the centres of
e.g., L1498 and L1517B (\cite{tafalla2002}, 2004). Similar result was found 
by Crapsi et al. (2007) using interferometric observations toward L1544.

\subsection{Ori B9}

Most of molecular material in the Orion complex is concentrated 
on the Orion A and B clouds. Star formation in Orion B (L1630) takes mainly
place in four clusters, NGC 2023, NGC 2024, NGC 2068, and NGC 2071 
(e.g., \cite{lada1992}; \cite{launhardt1996}).
The Orion B South cloud, which encompasses 
the star-forming regions NGC 2023/2024 is the only site of O and B star 
formation in Orion B (see, e.g., \cite{nutter2007}, hereafter NW07).
Apart of the above mentioned four regions, only single stars or small groups of
low- to intermediate mass stars are currently forming in Orion B 
(\cite{launhardt1996}).

Ori B9 lies in the central part of Orion B and is a relatively isolated cloud
at a projected distance of $\sim40\arcmin$ (5.2 pc at 450 pc\footnote{We assume
a distance to the Orion star-forming regions of 450 pc.}) northeast from
the closest star cluster NGC 2024 (\cite{caselli1995}) which is the most 
prominent region of current star formation in Orion B.
Ori B9 has avoided previous (sub)mm mappings which have
concentrated on the well-known active regions in the northern and
southern part of the GMC (NW07 and references therein).

In this paper we present results from the submm continuum
mapping of the Ori B9 cloud with LABOCA on APEX, and from spectral line 
observations towards three N$_2$H$^+$ peaks found by Caselli \& Myers
(1994) in the clump associated with the low-luminosity FIR source 
IRAS 05405-0117 
(see Fig.~2 in Caselli \& Myers (1994))\footnote{The N$_2$H$^+(1-0)$
map of Caselli and Myers (1994) shows two separated gas condensations of
$\sim0.1$ pc in size. The southern condensation has a weak subcomponent. 
The positions of our molecular line observations are given in Table~1 of 
Harju et al. (2006).}.
This source has the narrowest CS linewidth ($0.48$ km s$^{-1}$) in the Lada
et al. (1991) survey, and the narrowest NH$_3$ linewidth (average linewidth
is $0.29$ km s$^{-1}$) in the survey by Harju et al. (1993).
A kinetic temperature of 10 K was derived from ammonia in this region. 

We have previously detected the H$_2$D$^+$ ion towards two of the N$_2$H$^+$
peaks (\cite{harju2006}). These detections suggest that 1) the degree of
molecular depletion is high and 2) the ortho:para ratio of H$_2$ is low,
and thus the cores should have reached an evolved chemical stage.
The high density and low temperature may have resulted in CO
depletion. This possibility is supported by the fact that
the clump associated with IRAS 05405-0117 does not stand out in the CO map
of Caselli \& Myers (1995). 

In the present study we determine the properties and
spatial distribution of dense cores in the Ori B9 cloud. We also 
derive the degree of deuteration and ionization degree within the clump
associated with IRAS 05405-0117.

\vspace{1cm}

The observations and the data reduction procedures are described in Sect. 2. 
The observational results are presented in Sect. 3. In Sect. 4 we describe the 
methods used to derive the physical and chemical properties of the observed 
sources. In Sect. 5 we discuss the results of our study, and in Sect. 6
we summarise our main conclusions.

\section{Observations and data reduction}

\subsection{Molecular lines: N$_2$H$^+$ and N$_2$D$^+$}

The spectral line observations towards the three above mentioned N$_2$H$^+$
peaks were performed with the IRAM 30 m telescope on Pico Veleta, Spain, on
May 18--20, 2007.
The spectra were centred at the frequencies of the strongest N$_2$H$^+(1-0)$
and N$_2$D$^+(2-1)$ hyperfine components. We used the following rest
frequencies: 93173.777 MHz (N$_2$H$^+$($JF_1F=123\rightarrow012$), 
\cite{caselli1995b}) and 154217.154 MHz (N$_2$D$^+$($234\rightarrow123$),
\cite{gerin2001}). Dore et al. (2004), and very recently 
Pagani et al. (2009b), have refined the N$_2$H$^+$ and N$_2$D$^+$ line 
frequencies. The slight differences between the
"new" and "old" frequencies have, however, no practical effect on
to the radial velocities or other parameters derived here.
The observations were performed in the
frequency switching mode with the frequency throw set to 7.9 MHz for the 3 mm
lines and 15.8 MHz for the 2 mm lines.

As the spectral backend we used the VESPA
(Versatile Spectrometer Assembly) facility autocorrelator which has
a bandwidth of 20 MHz and a channel width of 10 kHz.
The lines were observed in two polarisations using the (AB)
100 GHz and the (CD) 150 GHz receivers. The horizontal polarisation 
at higher frequency (D150) turned out to be very noisy and was thus excluded
from the reduction.
The channel width used corresponds to 0.032 km s$^{-1}$ and 0.019 km s$^{-1}$
at the observed frequencies of N$_2$H$^+(1-0)$ and N$_2$D$^+(2-1)$,
respectively. The half-power beamwidth (HPBW) and the main beam efficiency,
$\eta_{\rm MB}$, are $26\farcs4$ and 0.80 at 93 GHz, and $16\arcsec$ and 0.73
at 154 GHz.

Calibration was achieved by the chopper wheel method. The pointing and focus 
were checked regularly towards Venus and several quasars.
Pointing accuracy is estimated to be better than $4-6\arcsec$.
The single-sideband (SSB) system temperatures were $\sim150-190$ K at 93 GHz
and $\sim290-340$ K at 154 GHz.
We reached an rms sensitivity in antenna temperature units of about
0.03 K in N$_2$H$^+(1-0)$ and about $0.05-0.07$ K in N$_2$D$^+(2-1)$.
The observational parameters are listed in Table~\ref{table:obs}. 

The CLASS programme, which is part of the GAG software developed at the
IRAM and the Observatoire de
Grenoble\footnote{{\tt http://www.iram.fr/IRAMFR/GILDAS}},
was used for the reductions.
Third order polynomial baselines were subtracted from the individual 
N$_2$H$^+(1-0)$ spectra before and after folding them.
From each individual N$_2$D$^+(2-1)$ spectra the
fourth order polynomial baselines were subtracted before folding.
Finally, the summed spectra were Hanning smoothed yielding the velocity 
resolutions of 0.064 km s$^{-1}$ for N$_2$H$^+(1-0)$ and 0.038 km s$^{-1}$ for 
N$_2$D$^+(2-1)$.
We fitted the lines using the hyperfine structure fitting method
of the CLASS programme. This method assumes that all the hyperfine components
have the same excitation temperature and width
($T_{\rm ex}$ and $\Delta {\rm v}$, respectively), 
and that their separations and relative line strengths are fixed to the values
given in Caselli et al. (1995), Gerin et al. (2001), and Daniel et al. (2006).
Besides $T_{\rm ex}$ and $\Delta {\rm v}$ this method gives an estimate of the
total optical depth, $\tau_{\rm tot}$, i.e, the sum of the peak optical depths
of the hyperfine components. These parameters can be used to estimate
the column density of the molecule.

\begin{table*}
\caption{Observational parameters.}
\begin{minipage}{2\columnwidth}
\centering
\renewcommand{\footnoterule}{}
\label{table:obs}
\begin{tabular}{c c c c c c c c c c c}
\hline\hline 
Molecule & Transition & Frequency & Instrument & $F_{\rm eff}$ & $\eta_{\rm MB}$\footnote{$\eta_{\rm MB}=B_{\rm eff}/F_{\rm eff}$, where $B_{\rm eff}$ and $F_{\rm eff}$ are the beam and forward efficiencies, respectively.} & \multicolumn{2}{c}{Resolution} & $F_{\rm throw}$ & $T_{\rm sys}$ & Obs. date\\
 & & [GHz] & & & & [\arcsec] & [km s$^{-1}$] & [MHz] & [K] & \\
\hline
N$_2$H$^+$ & $J=1-0$ & 93.173777\footnote{Frequency taken from Caselli et al. (1995).} & 30 m/AB100 & 0.95 & 0.80 & 26.4 & 0.064 & 7.9 & 150--190 & 18--20 May 2007\\
N$_2$D$^+$ & $J=2-1$ & 154.217154\footnote{Frequency taken from Gerin et al. (2001).} & 30 m/CD150 & 0.93 & 0.73 & 16 & 0.038 & 15.8 & 290--340 & 18--20 May 2007\\
\hline
\multicolumn{3}{c}{Continuum at 870 $\mu$m} & APEX/LABOCA & 0.97 & 0.73 & 18.6 & - & - & - & 4 Aug. 2007\\
\hline 
\end{tabular} 
\end{minipage}
\end{table*}

\subsection{Submillimetre continuum}

The 870 $\mu$m continuum observations toward the Ori B9 cloud were carried out
on 4 August 2007 with the 295 channel bolometer array LABOCA 
(Large APEX Bolometer Camera) on APEX.
The LABOCA central frequency is about 345 GHz and the bandwidth is about 
60 GHz. The HPBW of the telescope is $18\farcs6$ (0.04 pc at 450 pc)
at the frequency used. The total field of view (FoV) for the LABOCA is
$11\farcm4$. 
The telescope focus and pointing were checked using the planet Mars and
the quasar J0423-013.
The submm zenith opacity was determined using the sky-dip method and
the values varied from 0.16 to 0.20, with a median value of 0.18.
The uncertainty due to flux calibration is estimated to be $\sim10$\%.

The observations were done using the on-the-fly (OTF) mapping  mode with
a scanning speed of $3\arcmin$ s$^{-1}$. A single map consisted of 200 scans
of $30\arcmin$ in length in right ascension and spaced by $6\arcsec$ in
declination.
The area was observed three times, with a final sensitivity of about
0.03 Jy beam$^{-1}$ (0.1 M$_{\sun}$ beam$^{-1}$ assuming a dust temperature
of 10 K).

The data reduction was performed using the BoA (BOlometer Array Analysis
Software) software package according to guidelines in the BoA User and
Reference 
Manual (2007)\footnote{{\tt http://www.astro.uni-bonn.de/boawiki/Boa}}.
This included flat fielding, flagging bad/dark channels and data
according to telescope speed and acceleration, correcting for the atmospheric
opacity, division into subscans, baseline subtractions and median noise
removal, despiking, and filtering out the low frequencies of the $1/f$-noise.
Finally, the three individual maps were coadded. 

\subsection{Spitzer/MIPS archival data}

Pipeline (version S16.1.0) reduced ``post-BCD (Basic Calibrated Data)''
Spitzer/MIPS images at 24 and 70 $\mu$m were downloaded from the Spitzer
data archive using the Leopard
software package\footnote{{\tt http://ssc.spitzer.caltech.edu/propkit/spot/}}.

We used the software package MOPEX
(MOsaicker and Point source EXtractor)\footnote{
{\tt http://ssc.spitzer.caltech.edu/postbcd/mopex.html}} to perform aperture
and point-spread function (PSF) fitted photometry on the sources.
The point sources were extracted using the APEX package (distributed as part
of MOPEX).

At 24 $\mu$m a 5.31 pixel aperture with a sky annulus between 8.16 and 13.06
pixels for background subtraction were used. At 70 $\mu$m the pixel aperture 
was 8.75 pixels and the sky annulus ranged from 9.75 to 16.25 pixels.
The pixel scale is $2\farcs45$/pixel for 24 $\mu$m and $4\farcs0$/pixel
for 70 $\mu$m. The MIPS resolution is $\sim6\arcsec$ and $\sim18\arcsec$
at 24 and 70 $\mu$m, respectively. These values correspond to 0.01 pc and
0.04 pc at the cloud distance of 450 pc.
The aperture correction coefficients used with these settings
are 1.167 and 1.211 at 24 and 70 $\mu$m, respectively, as given on the Spitzer
Science Center (SSC) 
website\footnote{{\tt http://ssc.spitzer.caltech.edu/mips/apercorr}}.
The absolute calibration uncertainties are about 4\% for 24 $\mu$m, and about
10\% for 70 $\mu$m (\cite{engelbracht2007}; \cite{gordon2007}).

\section{Observational results}

\subsection{Dust emission}

\begin{figure*}
\centering
\includegraphics[width=18cm]{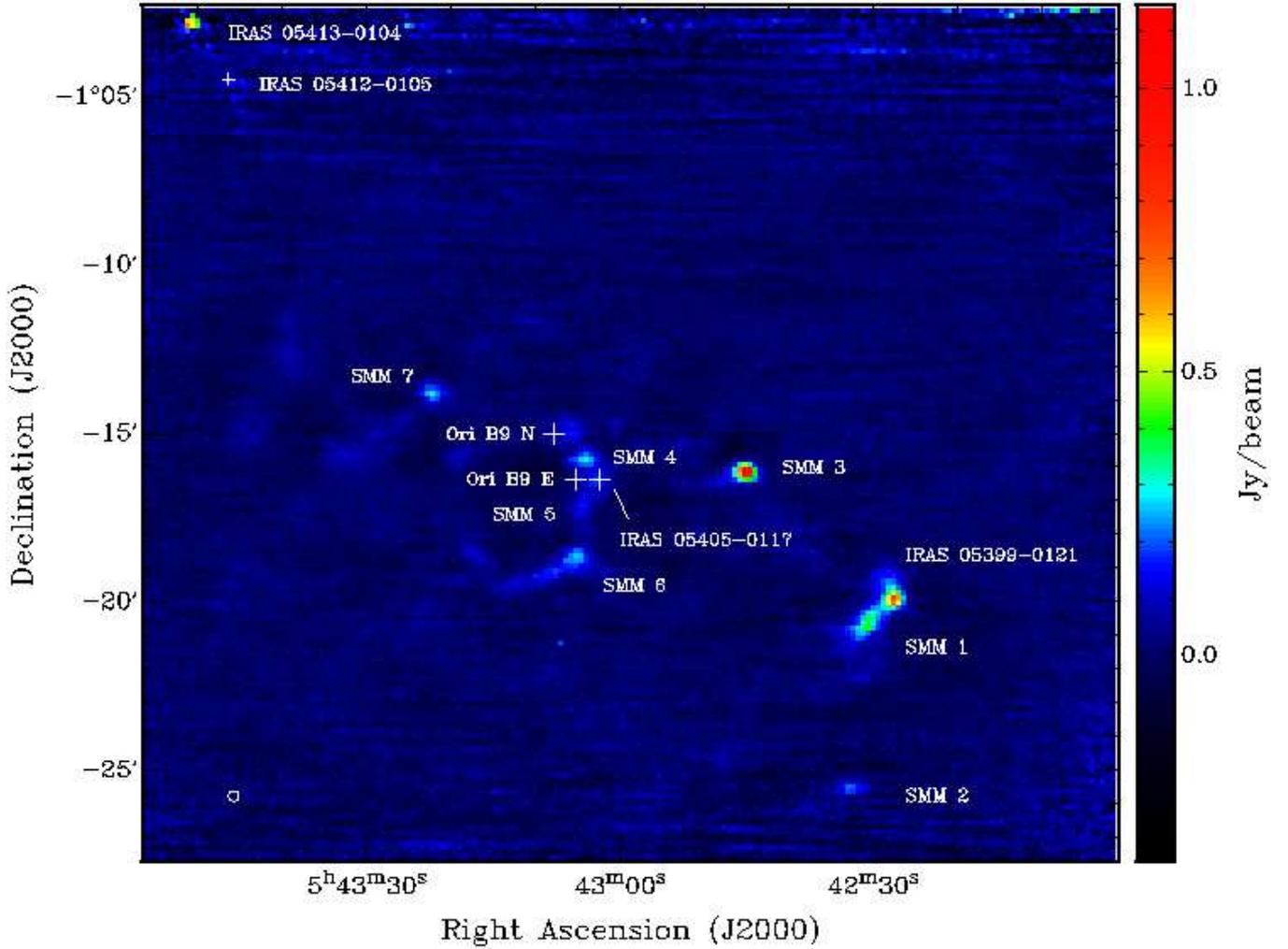}
\caption{LABOCA 870 $\mu$m map of the Ori B9 cloud. The three large
plus signs in the centre of the field mark the positions of our molecular line
observations (see Fig.~\ref{figure:positions}). The small plus sign in
the upper left shows the dust peak position of IRAS 05412-0105.
The beam HPBW ($18\farcs6$) is shown in the bottom left.}
\label{figure:map}
\end{figure*}

The obtained LABOCA map is presented in Fig.~\ref{figure:map}.
Altogether 12 compact sources can be identified on this map.
A source was deemed real if it had a peak flux density $>5\sigma$ 
(i.e., $>0.15$ Jy beam$^{-1}$) relative to the local background.
The coordinates, peak and integrated flux densities, deconvolved angular
FWHM diameters, and axis ratios of the detected sources are listed in 
Table~\ref{table:cores}.
The coordinates listed relate to the dust emission peaks. 
The integrated flux densities have been derived 
by summing pixel by pixel the flux density in the source area.
The uncertainty on flux density is derived from
$\sqrt{\sigma_{\rm cal}^2+\sigma_{\rm S}^2}$, where $\sigma_{\rm cal}$ is the
uncertainty from calibration, i.e., $\sim10$\% of flux density,
and $\sigma_{\rm S}$ is the uncertainty from flux density determination 
based on the rms noise level near the source area.
We have computed the deconvolved source angular diameters, $\theta_{\rm s}$, 
assuming that the brightness distribution is Gaussian.
The values of $\theta_{\rm s}$ correspond to the geometric mean
of the major and minor axes FWHM obtained from two-dimensional Gaussian fits to
the observed emission which has been corrected for the beam size. 
The uncertainty on $\theta_{\rm s}$ has been calculated by
propagating the uncertainties on the major and minor axes FWHM, which are
formal errors from the Gaussian fit. The axis ratio is defined as
the ratio of deconvolved major axis FWHM to minor axis FWHM.
Both the flux density determination and Gaussian fitting to the sources were 
done using the Miriad software package (\cite{sault1995}).

Four of the detected sources have IRAS (Infrared Astronomical Satellite) point
source counterparts, whereas eight are new submm sources.
We designate the eight new sources as SMM 1, SMM 2,
etc. The locations of three N$_2$H$^+(1-0)$ line emission peaks from Caselli
\& Myers (1994) are indicated on the map with plus signs 
(see Figs.~\ref{figure:map} and ~\ref{figure:positions}).
The N$_2$H$^+$ peak Ori B9 E which lies $40\arcsec$ east of IRAS 05405-0117
does not correspond to any submm peak (see Fig.~\ref{figure:positions}).
The N$_2$H$^+$ peak Ori B9 N lies about $39\arcsec$ southeast of the closest
dust continuum peak (see Sect. 5.5). One can see that our pointed
N$_2$H$^+$/ N$_2$D$^+$ observations, made before the LABOCA mapping, missed
the strongest submm peak SMM 4 located near IRAS 05405-0117.

\begin{table*}
\caption{Submillimetre sources in the Ori B9 cloud.}
\begin{minipage}{2\columnwidth}
\centering
\renewcommand{\footnoterule}{}
\label{table:cores}
\begin{tabular}{c c c c c c c}
\hline\hline 
 & \multicolumn{2}{c}{Peak position} & $S_{870}^{\rm peak}$ & $S_{870}$ & $\theta_{\rm s}$ & \\
Name & $\alpha_{2000.0}$ [h:m:s] & $\delta_{2000.0}$ [$\degr$:$\arcmin$:$\arcsec$] & [Jy beam$^{-1}$] & [Jy] & [\arcsec] & Axis ratio\\
\hline
IRAS 05399-0121 & 05 42 27.4 & -01 19 50 & 0.81 & $2.7\pm0.3$ & $30\pm4$ & 1.3\\
SMM 1 & 05 42 30.5 & -01 20 45 & 0.41 & $3.0\pm0.3$ & $57\pm7$ & 2.5\\
SMM 2 & 05 42 32.9 & -01 25 28 & 0.21 & $0.7\pm0.1$ & $26\pm5$ & 3.8\\
SMM 3 & 05 42 44.4 & -01 16 03 & 1.14 & $2.5\pm0.4$ & $19\pm3$ & 1.5\\
IRAS 05405-0117 & 05 43 02.7 & -01 16 21 & 0.19 & $0.9\pm0.1$\footnote{These values include both the IRAS 05405-0117 and Ori B9 E (see text and Fig.~\ref{figure:positions}).} & $45\pm8$ $^a$ & 1.2\\
SMM 4 & 05 43 03.9 & -01 15 44 & 0.28 & $1.3\pm0.1$ & $34\pm6$ & 2.3\\
SMM 5 & 05 43 04.5 & -01 17 06 & 0.16 & $0.6\pm0.1$ & $38\pm4$ & 1.8\\
SMM 6 & 05 43 05.1 & -01 18 38 & 0.26 & $2.5\pm0.3$ & $92\pm35$ & 4.1\\
Ori B9 N & 05 43 05.7 & -01 14 41 & 0.16 & $1.0\pm0.1$ & $47\pm5$ & 1.2\\
SMM 7 & 05 43 22.1 & -01 13 46 & 0.32 & $0.8\pm0.1$ & $24\pm4$ & 1.8\\
IRAS 05412-0105 & 05 43 46.4 & -01 04 30 & 0.17 & $0.5\pm0.1$ & - & -\\
IRAS 05413-0104 & 05 43 51.3 & -01 02 50 & 0.66 & $0.9\pm0.2$ & $25\pm8$ & 1.6\\
\hline 
\end{tabular} 
\end{minipage}
\end{table*}

\subsection{Spitzer/MIPS images}

\begin{figure}[!h]
\resizebox{\hsize}{!}{\includegraphics[angle=270]{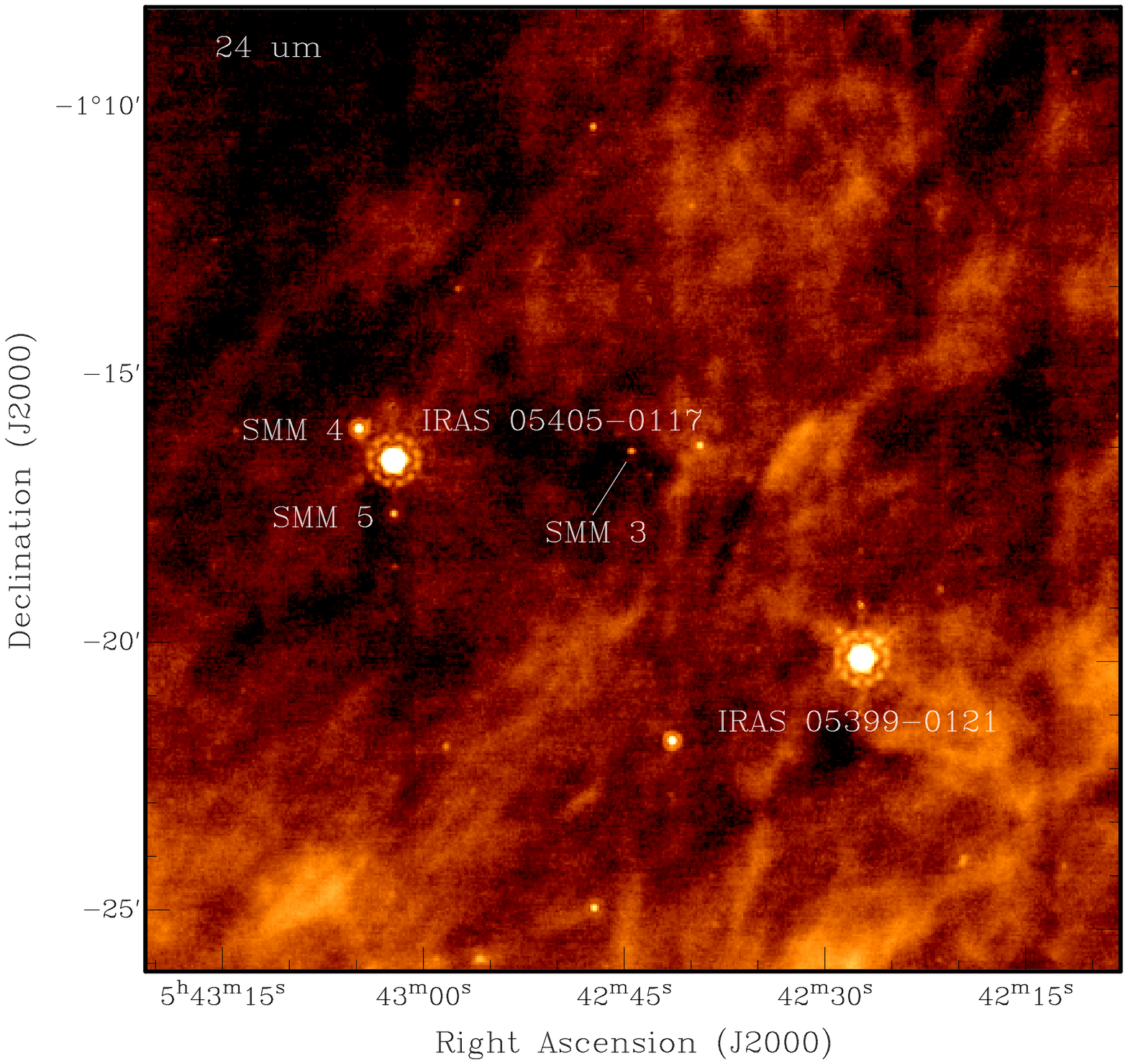}}
\resizebox{\hsize}{!}{\includegraphics{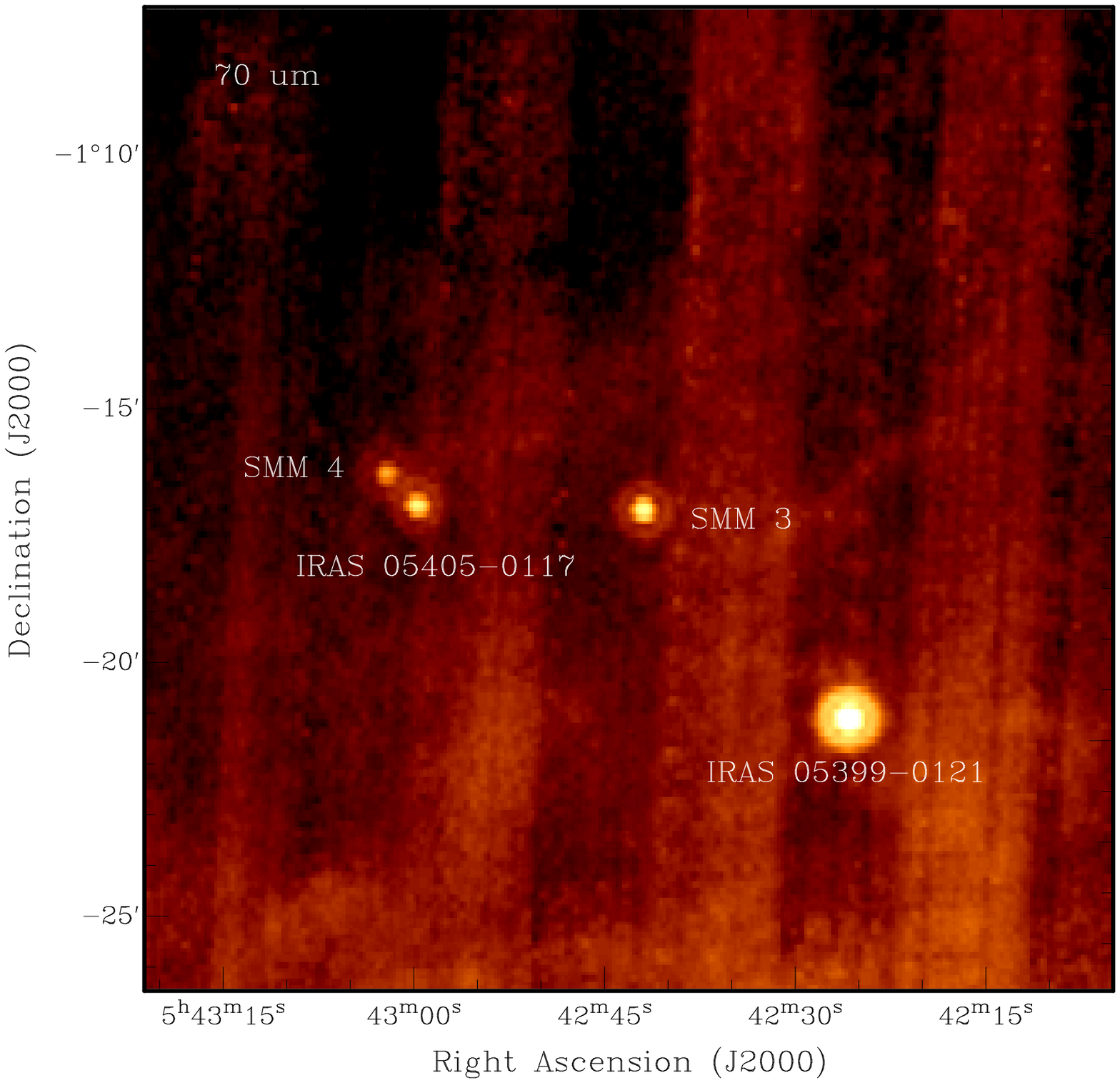}}
\caption{Spitzer/MIPS 24 $\mu$m (top) and 70 $\mu$m (bottom) images of the 
central part of the Ori B9 cloud. The logarithmic colour scale range
from 33.4 to 729.5 MJy sr$^{-1}$ and 27.9 to 1357.4 MJy sr$^{-1}$ in the 24 
and 70 $\mu$m images, respectively.}
\label{figure:spitzer}
\end{figure}

The retrieved Spitzer/MIPS images at 24 and 70 $\mu$m are presented in
Fig.~\ref{figure:spitzer}. All four IRAS sources that were detected by LABOCA
are visible at both 24 and 70 $\mu$m (IRAS 05412-0105 and 05413-0104 northeast
from the central region are outside the regions shown in
Fig.~\ref{figure:spitzer}). From the new submm sources, SMM 3 and
SMM 4 are also visible at both 24 and 70 $\mu$m, while there is 24 $\mu$m
source near SMM 5 which is not detected at 70 $\mu$m.
The rest of the submm sources are visible at neither of the wavelenghts.

In Table~\ref{table:spitzer} we list the sources detected at both 24 and 70
$\mu$m. In this table we give the 24 and 70 $\mu$m peak positions of the
sources and their flux densities at both wavelengths obtained from the
aperture photometry. The $1\sigma$ uncertainties on flux densities were
derived as described in Sect. 3.1, i.e. as a quadratic sum of the
calibration and photometric uncertainties.

\begin{table*}
\caption{Spitzer 24/70 $\mu$m sources in Ori B9.}
\begin{minipage}{2\columnwidth}
\centering
\renewcommand{\footnoterule}{}
\label{table:spitzer}
\begin{tabular}{c c c c c c c c c}
\hline\hline 
 & \multicolumn{2}{c}{24 $\mu$m peak position} & \multicolumn{2}{c}{70 $\mu$m peak position} & $S_{24}$ & $S_{70}$ \\
Name & $\alpha_{2000.0}$ [h:m:s] & $\delta_{2000.0}$ [$\degr$:$\arcmin$:$\arcsec$] & $\alpha_{2000.0}$ [h:m:s] & $\delta_{2000.0}$ [$\degr$:$\arcmin$:$\arcsec$] & [Jy] & [Jy]\\
\hline
IRAS 05399-0121 & 05 42 27.6 & -01 20 01 & 05 42 27.7 & -01 19 57 & $1.3\pm0.05$ & $24.4\pm2.4$\\
SMM 3 & 05 42 45.3 & -01 16 14 & 05 42 45.1 & -01 16 13 & $0.005\pm0.0002$ & $3.6\pm0.4$\\
IRAS 05405-0117 & 05 43 03.1 & -01 16 29 & 05 43 03.0 & -01 16 30 & $1.3\pm0.05$ & $2.7\pm0.3$\\
SMM 4 & 05 43 05.7 & -01 15 55 & 05 43 05.6 & -01 15 52 & $0.036\pm0.001$ & $1.3\pm0.1$\\
IRAS 05412-0105 & 05 43 46.3 & -01 04 44 & 05 43 46.1 & -01 04 43 & $0.6\pm0.02$ & $2.5\pm0.3$\\
IRAS 05413-0104 & 05 43 51.4 & -01 02 53 & 05 43 51.3 & -01 02 51 & $0.2\pm0.01$ & $17.9\pm1.8$\\
\hline 
\end{tabular} 
\end{minipage}
\end{table*}

\subsection{N$_2$H$^+$ and N$_2$D$^+$}

\begin{figure}[!h]
\resizebox{\hsize}{!}{\includegraphics[angle=270]{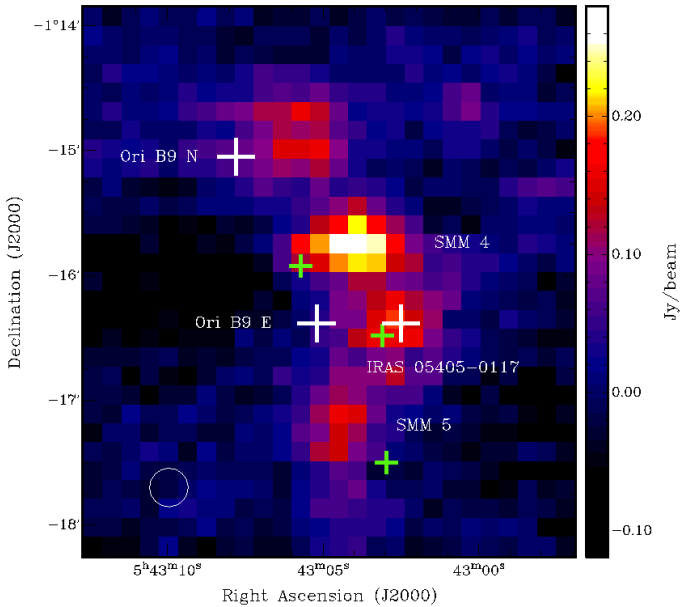}}
\caption{Blow-up of Fig.~\ref{figure:map} showing the IRAS 05405-0117 clump
region. The large plus signs indicate the positions of our H$_2$D$^+$,
N$_2$H$^+$, and N$_2$D$^+$ observations towards three condensations shown in
Fig.~2 of Caselli \& Myers (1994). Also shown are the 24 $\mu$m peak positions
of SMM 4 and IRAS 05405-0117, and the 24 $\mu$m peak near SMM 5 (small green
plus signs, cf. Fig.~\ref{figure:spitzer}). The beam HPBW ($18\farcs6$) is
shown in the bottom left.}
\label{figure:positions}
\end{figure}

The three positions of our molecular line observations are indicated
in Figs.~\ref{figure:map} and ~\ref{figure:positions}.
These positions correspond to the three
N$_2$H$^+(1-0)$ peaks found by Caselli \& Myers (1994) (see also Sect. 1.1).
The Hanning smoothed N$_2$H$^+(1-0)$ and N$_2$D$^+(2-1)$ spectra are
shown in Figs.~\ref{figure:n2h+} and \ref{figure:n2d+}, respectively.
The seven hyperfine components of N$_2$H$^+(1-0)$ are clearly resolved
towards all three positions. The N$_2$H$^+(1-0)$ spectra
towards Ori B9 E and Ori B9 N show additional lines, which can be explained 
by N$_2$H$^+(1-0)$ emission originating at a different radial velocity 
(see Fig.~\ref{figure:n2h+}). In the case of Ori B9 E, the additional 
N$_2$H$^+(1-0)$ lines originate in gas at a radial velocity of 1.3 km s$^{-1}$,
whereas towards Ori B9 N the additional gas component has a ${\rm v}_{\rm LSR}$
of 2.2 km s$^{-1}$. These velocities are $\sim7-8$ km s$^{-1}$ lower than the
average velocity of the Ori B9 cloud, suggesting that they are produced by 
a totally different gas component.
We checked that the additional components are not caused by, e.g., a 
phase-lock failure by summing randomly selected subsets of the spectra. 
All sums constructed in this manner showed the same features with
equal intensity ratios. The ``absorption''-like
feature at $\sim20$ km s$^{-1}$ in the N$_2$H$^+(1-0)$ spectrum of Ori B9 N
is an arfefact caused by the frequency switching folding process.

Only the strongest hyperfine group of N$_2$D$^+(2-1)$ was detected.
The relatively poor S/N ratio hampers the hyperfine component
fitting. Towards Ori B9 N, the additional velocity component at 
$\sim2.2$ km s$^{-1}$ was also detected in N$_2$D$^+(2-1)$.

In Table~\ref{table:line_parameters} we give the N$_2$H$^+(1-0)$ and
N$_2$D$^+(2-1)$ line parameters derived from Hanning smoothed spectra.
The LSR velocities and linewidths (FWHM) are
listed in columns (2) and (3), respectively. The total optical depth and
excitation temperature for the lines are given in columns (6) and (7),
respectively. The excitation temperatures, $T_{\rm ex}$, of the
N$_2$H$^+(1-0)$ transition were derived from the antenna equation

\begin{equation}
\label{eq:ant_1}
T_{\rm A}^{*}= \eta \frac{h\nu}{k_{\rm B}}\left[F(T_{\rm ex})-F(T_{\rm bg})\right]\left(1-e^{-\tau}\right) \, ,
\end{equation}
where $\eta$ is the beam-source coupling efficiency, $h$ is the 
Planck constant, $\nu$ is the transition frequency, $k_{\rm B}$ is the 
Boltzmann constant, $T_{\rm bg}=2.725$ K is the cosmic microwave background
(CMB) temperature, and the function $F(T)$ is defined by 
$F(T)\equiv\left(e^{h\nu/k_{\rm B}T}-1\right)^{-1}$.
We assumed that $\eta=\eta_{\rm MB}$, and we used the main beam 
brightness temperature, $T_{\rm MB}=\eta_{\rm MB}^{-1}T_{\rm A}^*$, 
and the optical thickness, $\tau$, of the brightest hyperfine component.
The uncertainty of $T_{\rm ex}$ has been calculated by propagating 
the uncertainties on $T_{\rm MB}$ and $\tau$.

The total optical depth of the N$_2$D$^+(2-1)$ line cannot be calculated
directly because the hyperfine components are not resolved in the spectra. 
We estimated the total optical depth in the following manner. First, we
calculated the optical depth of the main hyperfine group of N$_2$D$^+(2-1)$
from the antenna equation using the $T_{\rm MB}$ obtained from a Gaussian fit
to the group of 4 strongest hyperfines. In the calculation, we
adopted the excitation temperature of the N$_2$H$^+(1-0)$ lines.
Second, the total optical depths of N$_2$D$^+(2-1)$ were
calculated taking into account that the main group correspond to 54.3\% of
$\tau_{\rm tot}$.
The uncertainty on $\tau_{\rm tot}$ has been calculated by propagating 
the uncertainties on $T_{\rm MB}$ and $T_{\rm ex}$.

\begin{table*}
\begin{minipage}{2\columnwidth}
\caption{N$_2$H$^+(1-0)$ and N$_2$D$^+(2-1)$ line parameters derived from
Hanning smoothed spectra. The integrated line intensity
($\int T_{\rm A}^*({\rm v}){\rm dv}$) includes all the
hyperfine components in the case of N$_2$H$^+$, whereas for N$_2$D$^+$ only
the main group is included.}
\centering
\renewcommand{\footnoterule}{}
\label{table:line_parameters}
\begin{tabular}{c c c c c c c}
\hline\hline 
Line/Position & ${\rm v}_{\rm LSR}$ [km s$^{-1}$] & $\Delta {\rm v}$ [km s$^{-1}$] & $T_{\rm A}^*$ [K] & $\int T_{\rm A}^*({\rm v}){\rm dv}$ [K km s$^{-1}$] & $\tau_{\rm tot}$ & $T_{\rm ex}$ [K]\\
\hline
{\bf N$_2$H$^+$}$(1-0)$\\
IRAS 05405-0117\footnote{Caselli \& Myers (1994) derived ${\rm v}_{\rm LSR}=9.209\pm0.003$ km s$^{-1}$, $\Delta {\rm v}=0.313\pm0.008$ km s$^{-1}$, and $\tau_{\rm tot}=4.594\pm0.825$.} & $9.228\pm0.001$ & $0.290\pm0.002$ & $2.37\pm0.04$ & $3.97\pm0.05$ & $6.1\pm0.03$ & $6.8\pm0.07$\\ 
Ori B9 E\footnote{For the other velocity component hyperfine fit yields ${\rm v}_{\rm LSR}=1.310\pm0.013$ km s$^{-1}$, $\Delta {\rm v}=0.436\pm0.035$ km s$^{-1}$, and $\tau_{\rm tot}=6.6\pm2.0$.} & $9.163\pm0.002$ & $0.298\pm0.005$ & $1.46\pm0.04$ & $2.26\pm0.03$ & $3.5\pm0.5$ & $6.1\pm0.3$\\ 
Ori B9 N\footnote{For the other velocity component ${\rm v}_{\rm LSR}=2.219\pm0.006$ km s$^{-1}$, $\Delta {\rm v}=0.378\pm0.021$ km s$^{-1}$, $T_{\rm A}^*=0.52\pm0.03$ K, $\tau_{\rm tot}=1.0$, and $T_{\rm ex}=5.9\pm0.17$ K.} & $9.149\pm0.003$ & $0.261\pm0.008$ & $1.82\pm0.09$ & $2.29\pm0.04$ & $2.2\pm0.8$ & $8.3\pm1.4$\\ 
{\bf N$_2$D$^+$}$(2-1)$\\
IRAS 05405-0117 & $9.414\pm0.012$ & $0.319\pm0.027$ & $0.31\pm0.04$ & $0.13\pm0.01$ & $0.26\pm0.01$\footnote{$\tau_{\rm tot}$ calculated by taking into account
that the main hyperfine group corresponds to 54.3\% of the total optical
depth. $\tau_{\rm main \, group}$ is calculated using $T_{\rm MB}$ from Gaussian fit to the main group and $T_{\rm ex}$ from N$_2$H$^+(1-0)$.} & $6.8\pm0.07$\\Ori B9 E & $9.285\pm0.013$ & $0.194\pm0.029$ & $0.27\pm0.03$ & $0.09\pm0.01$ & $0.28\pm0.03^d$ & $6.1\pm0.3$\\ 
Ori B9 N\footnote{For the other velocity component ${\rm v}_{\rm LSR}=2.290\pm0.044$ km s$^{-1}$, $\Delta {\rm v}=0.187\pm0.090$ km s$^{-1}$, $T_{\rm A}^*=0.22\pm0.06$ K, and $\tau_{\rm tot}=0.12\pm0.01$ ($\tau_{\rm tot}$ is calculated as described in footnote $d$).} & $9.255\pm0.009$ & $0.136\pm0.021$ & $0.28\pm0.07$ & $0.09\pm0.01$ & $0.16\pm0.05^d$ & $8.3\pm1.4$\\ 
\hline 
\end{tabular} 
\end{minipage} 
\end{table*}

\section{Physical and chemical parameters of the sources}

\subsection{Spectral energy distributions}

The 24 and 70 $\mu$m flux densities together with the integrated flux
densities at 870 $\mu$m were used to fit the spectral energy
distribution (SED) of SMM 3 and SMM 4. For the IRAS sources also the archival
IRAS data were included.
The flux densities in the 12, 25, 60, and 
100 $\mu$m IRAS bands are listed in Table~\ref{table:iras}.
The derived SEDs for SMM 3, SMM 4, and IRAS 05405-0117 are shown in
Fig.~\ref{figure:sed}.

For all six sources detected at three or more wavelengths, the data were
fitted with a two-temperature composite model.
The parameters resulting from the fitting are given in Table~\ref{table:sed}.
We adopted a gas-to-dust mass ratio of 100 and dust opacities  
corresponding to a MRN size distribution with thick ice mantles at a gas
density of $n_{\rm H}=10^5$ cm$^{-3}$ (\cite{ossenkopf1994}).
The total (cold+warm) mass and the integrated bolometric luminosity are given
in columns (2) and (3) of Table~\ref{table:sed}, respectively.
The temperatures of the two components are listed
in columns (4) and (5). In columns (6) and (7) we give the mass and luminosity
fractions of the cold component vs. the total mass and luminosity, and in
column (8) we list the ratio of submm luminosity
(numerically integrated longward of 350 $\mu$m) to total bolometric luminosity
($L_{\rm submm}/L_{\rm bol}$). Column (9) list the normalised
envelope mass, $M_{\rm tot}/L_{\rm bol}^{0.6}$, which is an
evolutionary indicator in the sense that it correlates with the protostellar 
outflow strength (i.e., with the mass accretion rate), and thus 
decreases with time (\cite{bontemps1996}).
In column (10) we give the source SED classification (see Sect. 5.1). 
In all cases the mass of the warm component is negligible 
($\sim10^{-7}-10^{-4}$ M$_{\sun}$) and thus the bulk of the material is
cold ($M_{\rm cold}/M_{\rm tot}\sim1$). 

\begin{table}
\caption{IRAS flux densities in Jy.}
\centering
\renewcommand{\footnoterule}{}
\label{table:iras}
\begin{tabular}{c c c c c}
\hline\hline 
Name & $S_{12}$ & $S_{25}$ & $S_{60}$ & $S_{100}$\\
\hline
IRAS 05399-0121 & 0.25 & 1.59 & 22.94 & 45.93\\
IRAS 05405-0117 & 0.40 & 1.55 & 3.75 & 19.67\\
IRAS 05412-0105 & 0.26 & 0.65 & 1.66 & 73.58\\
IRAS 05413-0104 & 0.25 & 0.31 & 17.33 & 59.46\\
\hline 
\end{tabular} 
\end{table}

\begin{figure}[!h]
\resizebox{\hsize}{!}{\includegraphics[angle=270]{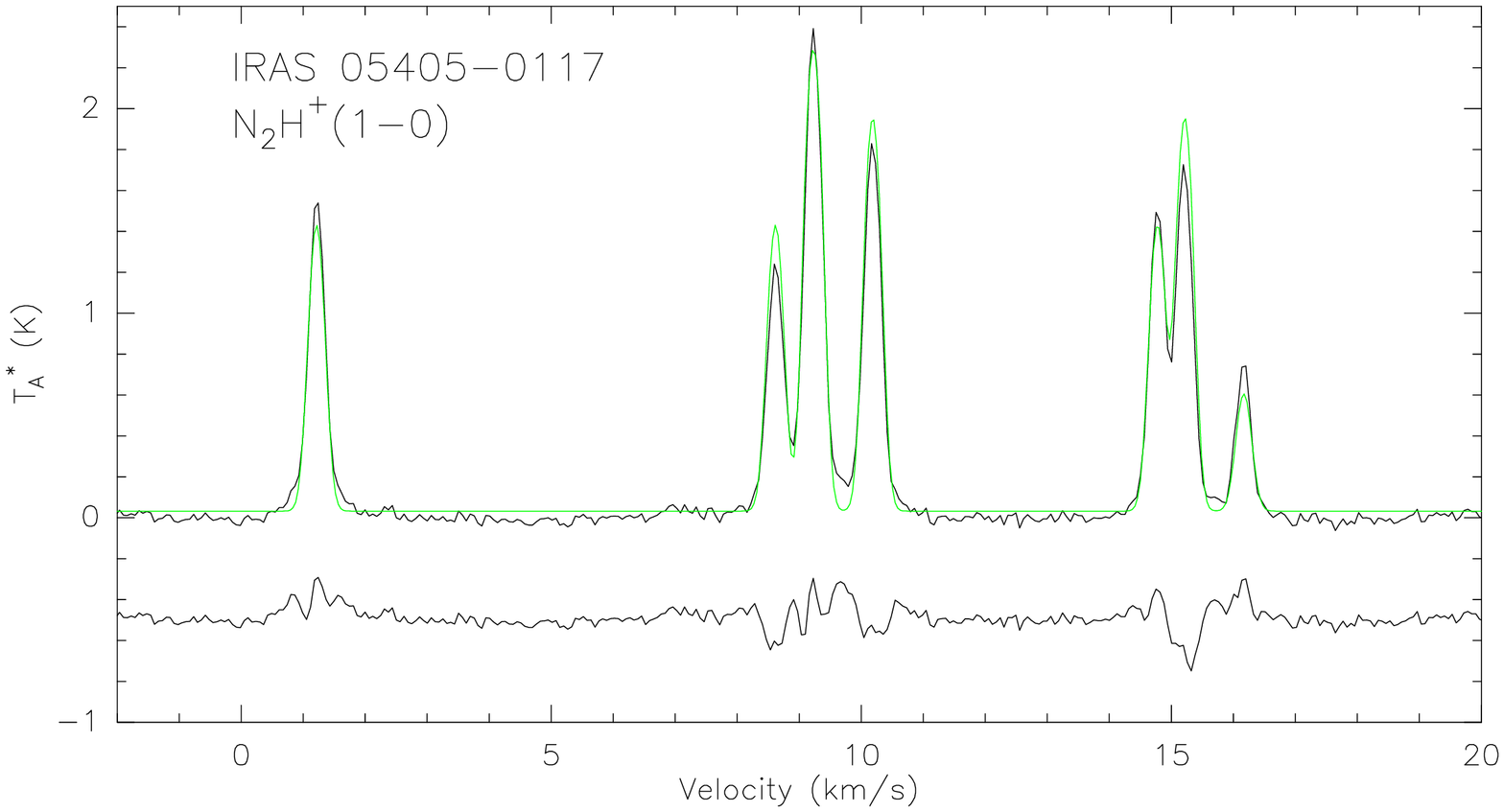}}
\resizebox{\hsize}{!}{\includegraphics[angle=270]{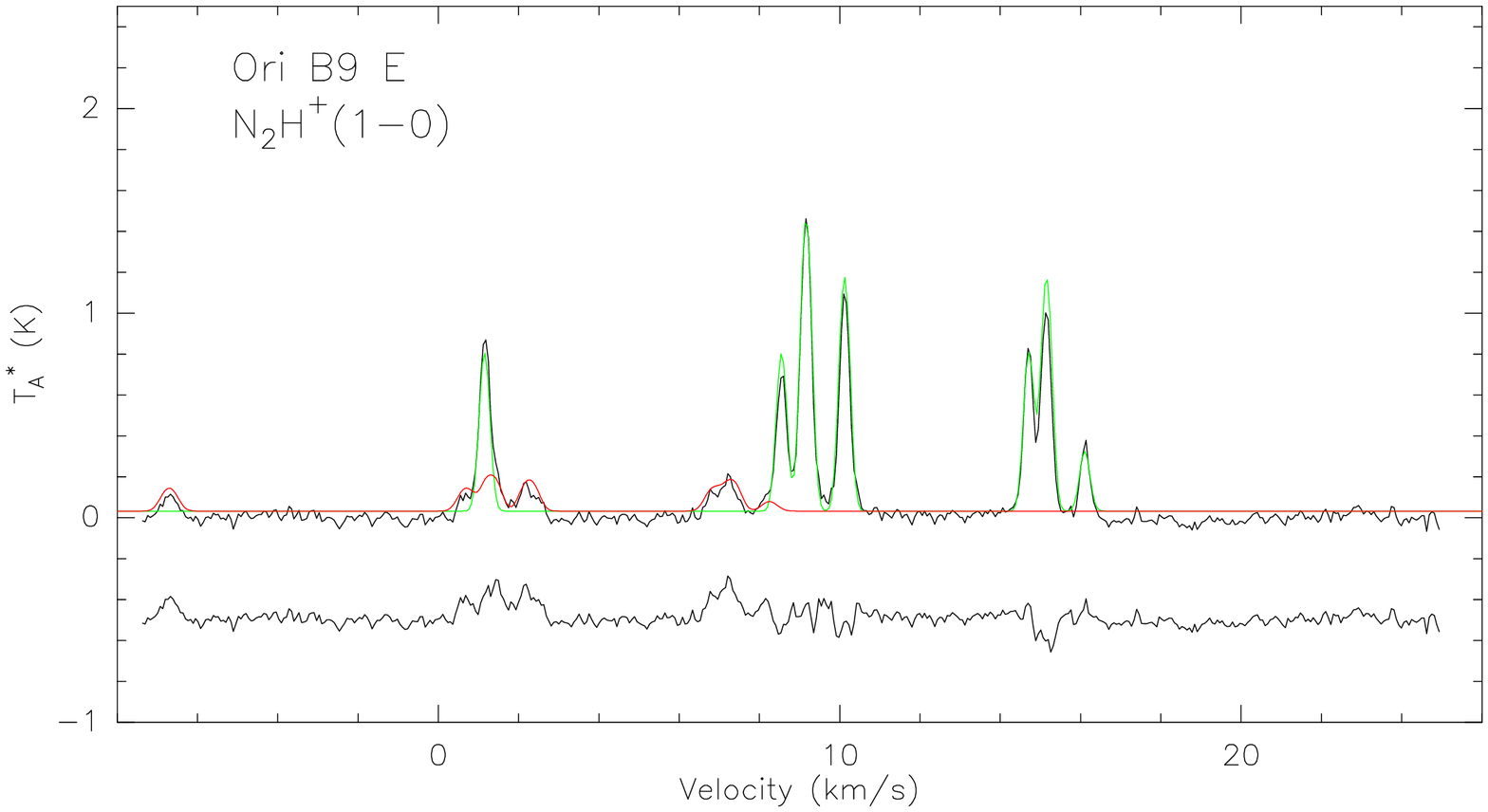}}
\resizebox{\hsize}{!}{\includegraphics[angle=270]{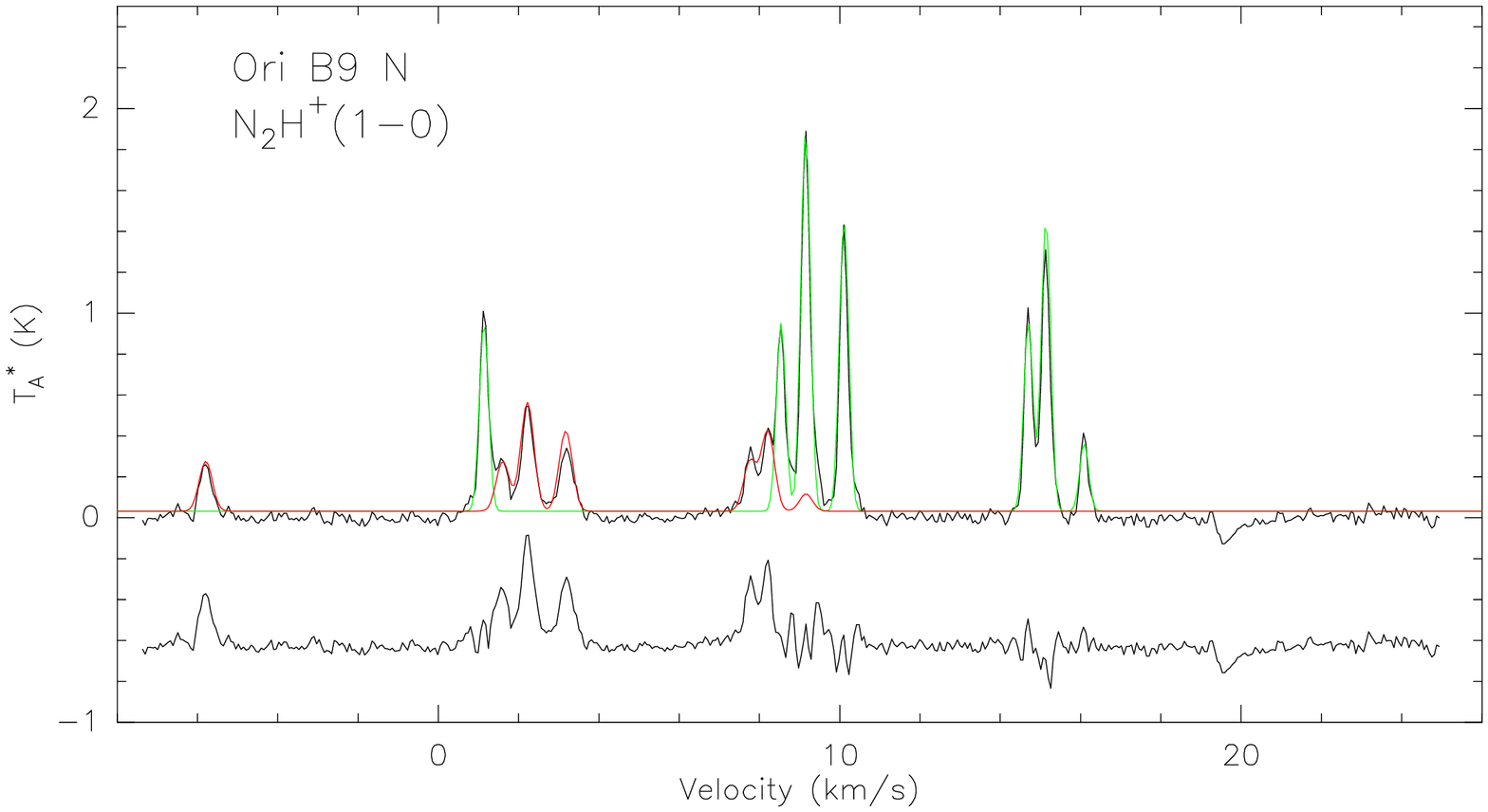}}
\caption{N$_2$H$^+(1-0)$ spectra toward IRAS 05405-0117 (top), Ori B9 E 
(middle), and Ori B9 N (bottom) after Hanning smoothing.
Hyperfine fits to the spectra are indicated with green lines.
The residuals of the fits are shown below the spectra. Hyperfine fits to the
other velocity component are indicated with red lines (see text). The
small ``absorption''-like feature at $\sim20$ km s$^{-1}$ in the bottom panel
is arfefact caused by frequency switching.}
\label{figure:n2h+}
\end{figure}

\begin{figure}[!h]
\resizebox{\hsize}{!}{\includegraphics[angle=270]{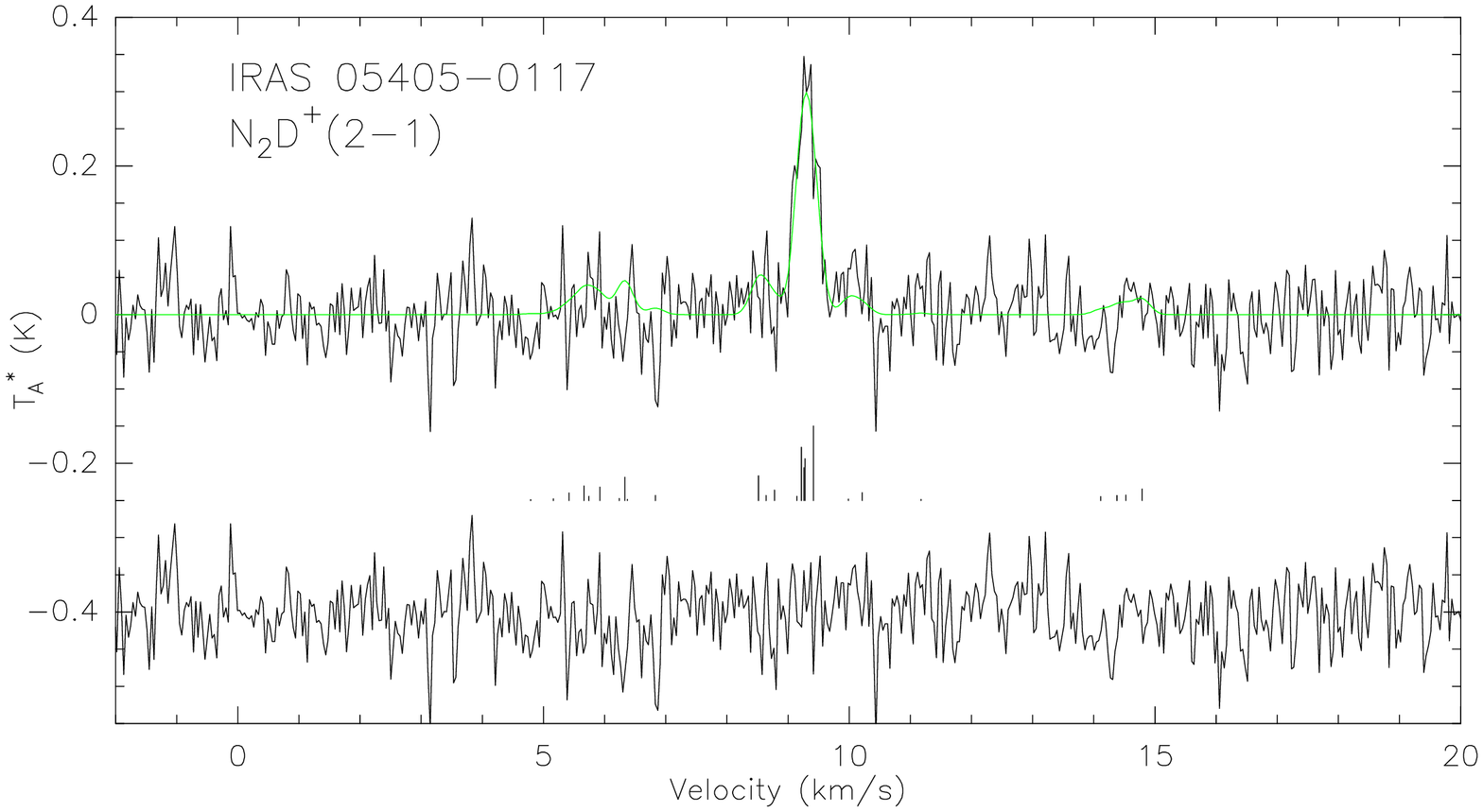}}
\resizebox{\hsize}{!}{\includegraphics[angle=270]{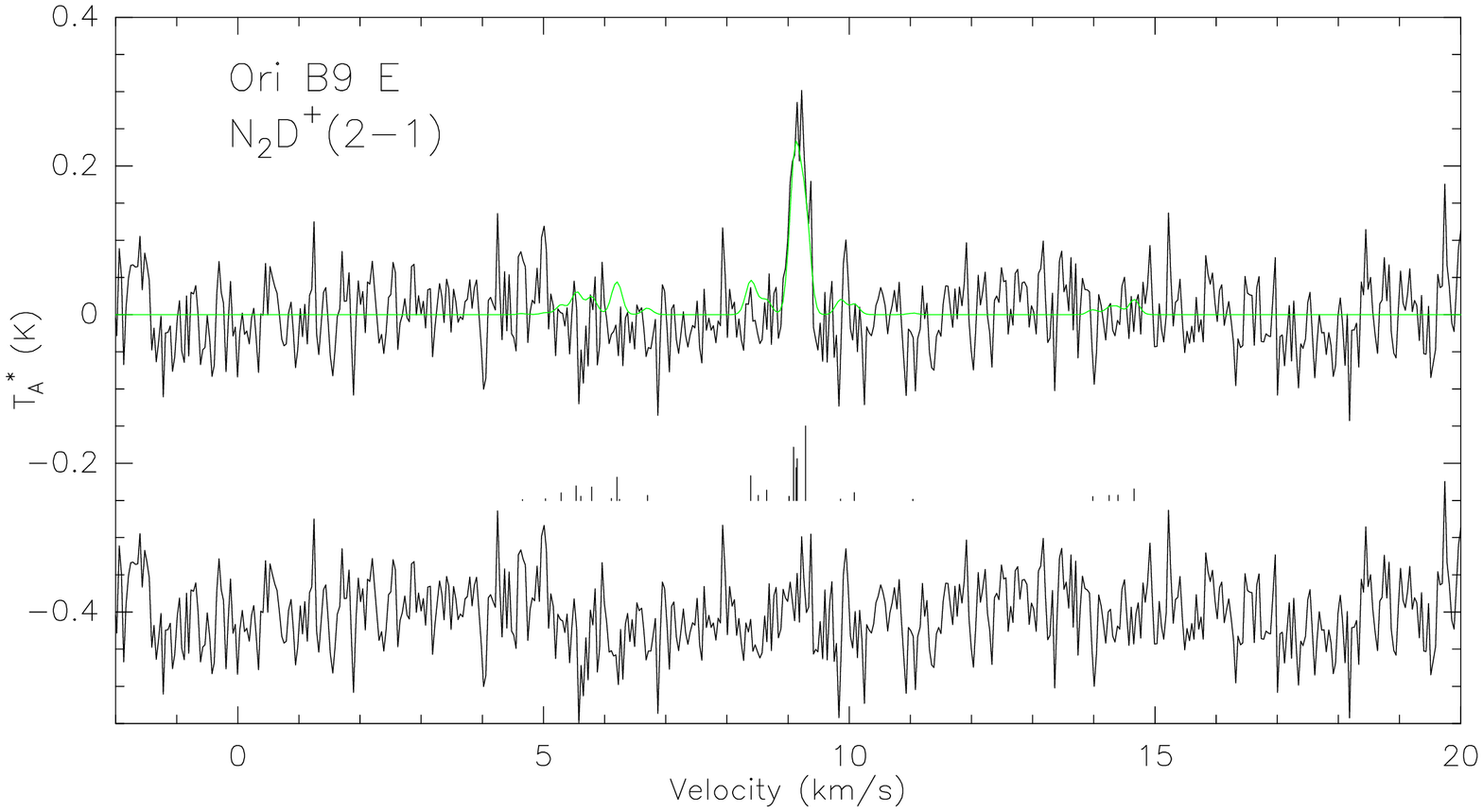}}
\resizebox{\hsize}{!}{\includegraphics[angle=270]{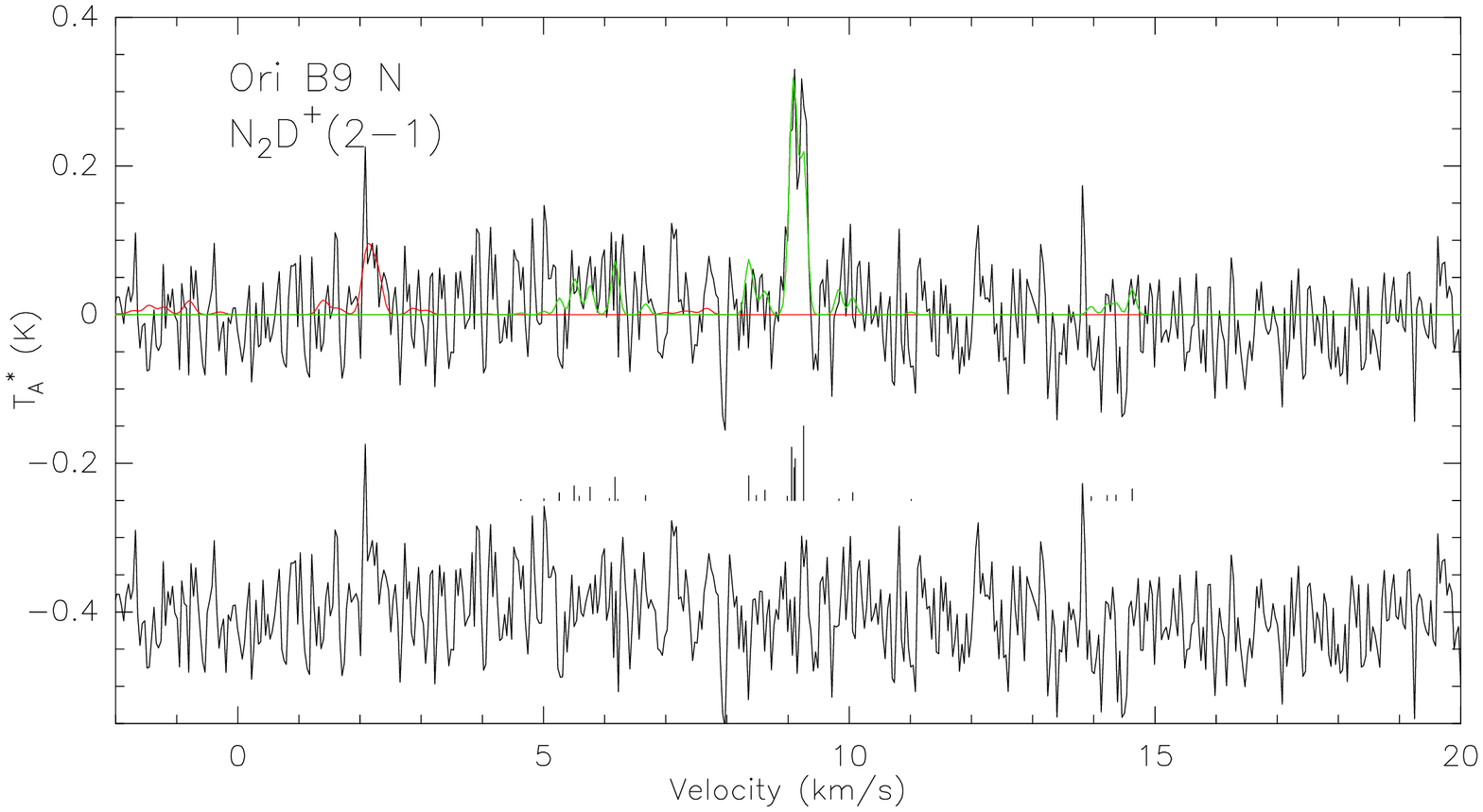}}
\caption{N$_2$D$^+(2-1)$ spectra toward IRAS 05405-0117 (top), Ori B9 E
(middle), and Ori B9 N (bottom) after Hanning smoothing. Hyperfine fits to the
spectra are indicated with green lines. The lines under the spectra indicate
the positions and relative intensities of the hyperfine components
(see Table~2 in Gerin et al. (2001)). Undermost are plotted the residuals of
the fits. Hyperfine fit to the other velocity component in the bottom
panel is indicated with red line (see text).}
\label{figure:n2d+}
\end{figure}

\begin{figure}[!h]
\resizebox{\hsize}{!}{\includegraphics{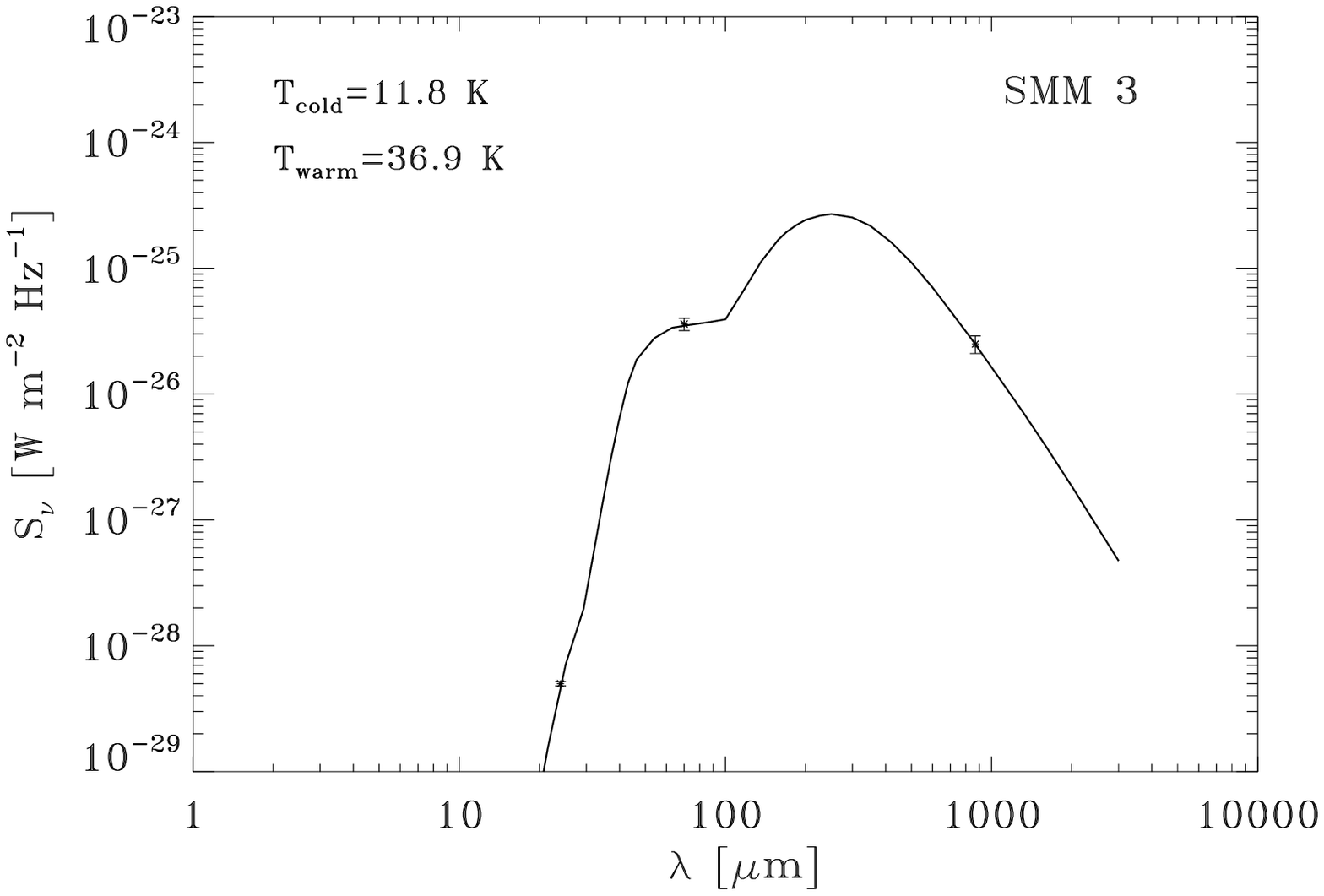}}
\resizebox{\hsize}{!}{\includegraphics{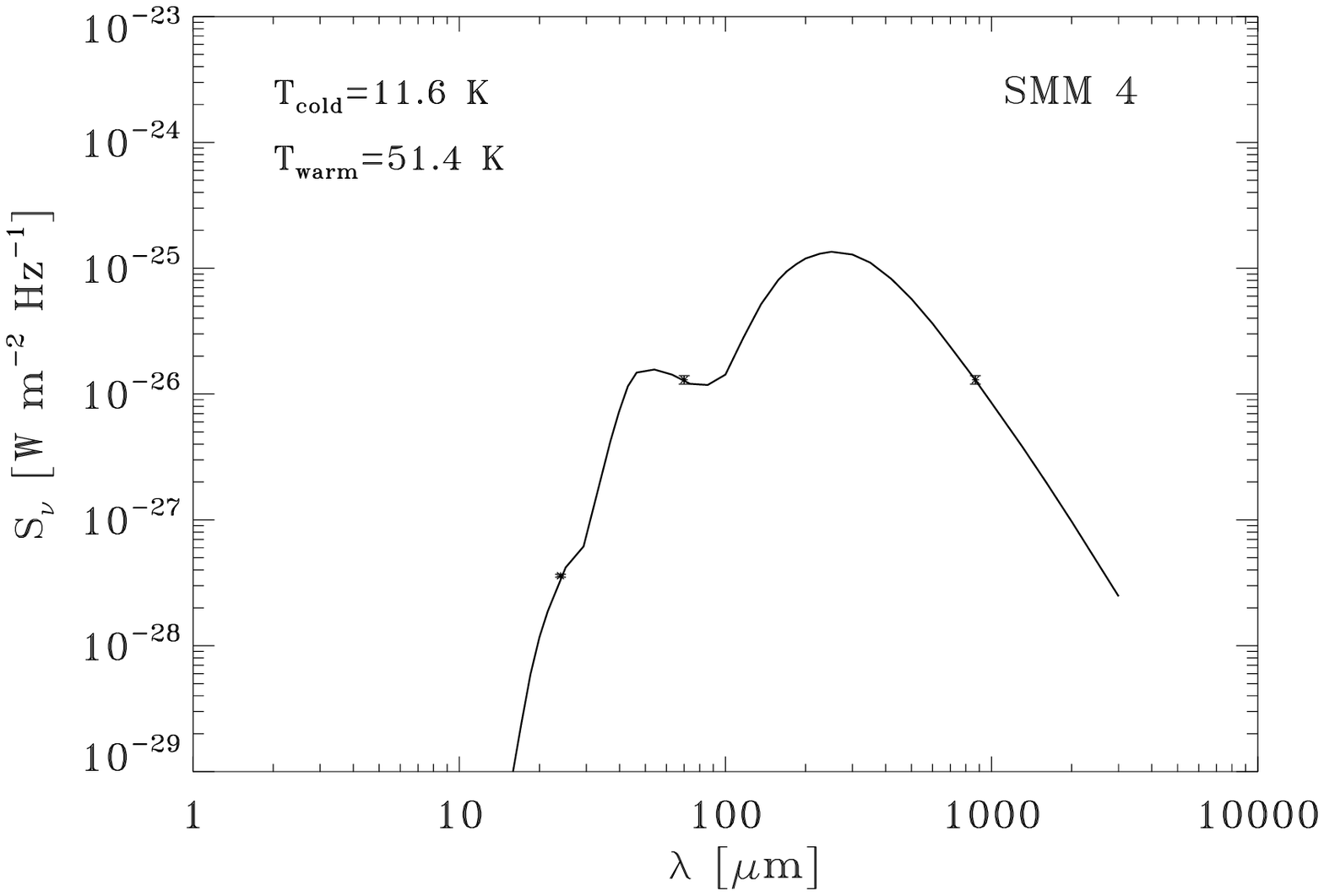}}
\resizebox{\hsize}{!}{\includegraphics{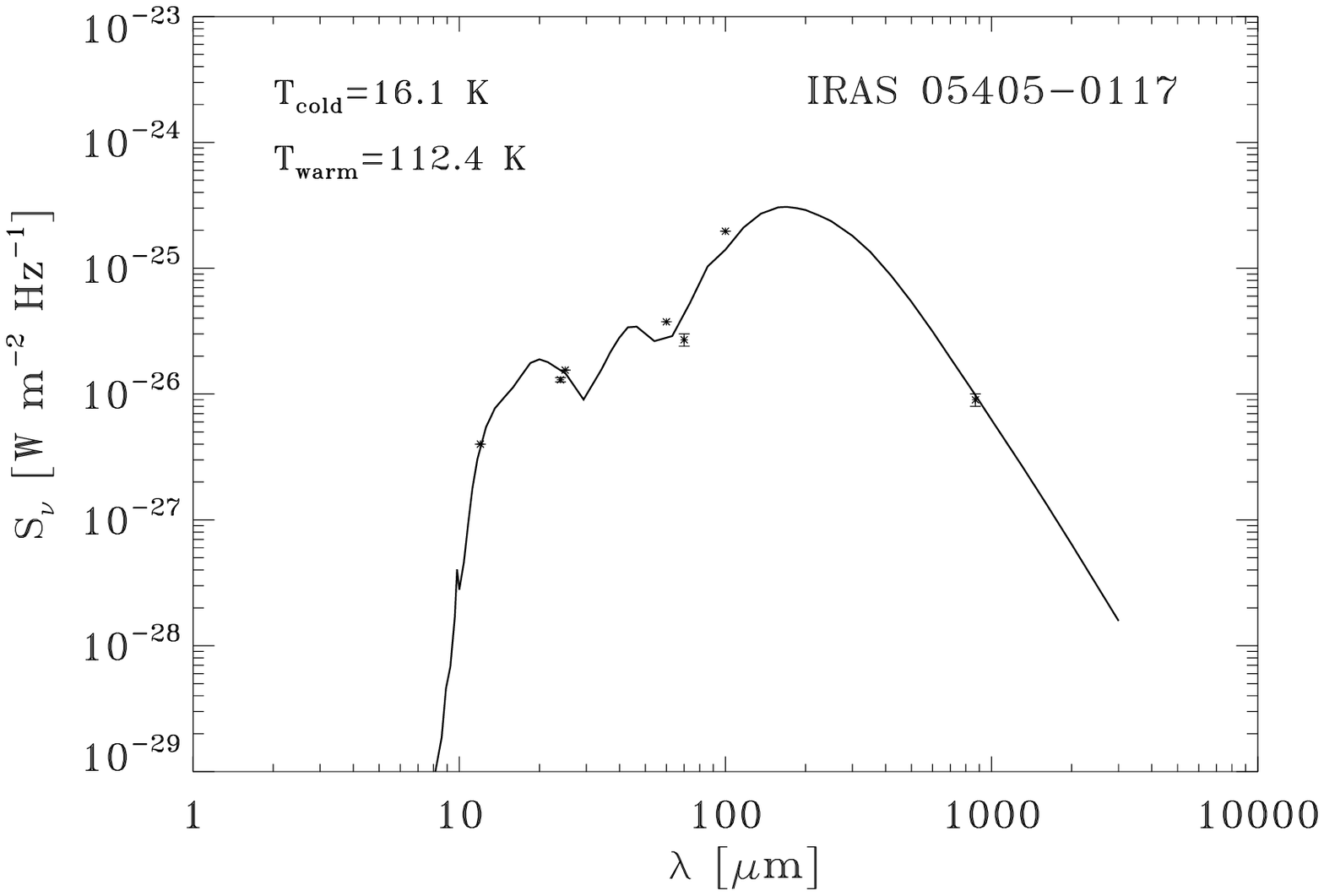}}
\caption{SEDs of the sources SMM 3 (top), SMM 4 (middle), and IRAS 05045-0117
(bottom). 24 and 70 $\mu$m data points are derived from archival Spitzer/MIPS
data, while the 870 $\mu$m measurement is performed with LABOCA.
For IRAS 05045-0117 we include also IRAS archival data 
(see Table~\ref{table:iras}). $1\sigma$ error bars are indicated for Spitzer 
and LABOCA data points. The solid lines in all plots represent the sum
of two (cold$+$warm) components (see columns (4) and (5) of 
Table~\ref{table:sed}).
The ``ripple'' between $\sim30-60$ $\mu$m in the SED of IRAS 05045-0117 is due
to simple logarithmic interpolation used to derive the luminosity.}
\label{figure:sed}
\end{figure}

\begin{table*}
\caption{Results of the SED fits.}
\begin{minipage}{2\columnwidth}
\centering
\renewcommand{\footnoterule}{}
\label{table:sed}
\begin{tabular}{c c c c c c c c c c c}
\hline\hline 
    & $M_{\rm tot}$ & $L_{\rm bol}$ & $T_{\rm cold}$ & $T_{\rm warm}$ &  & &  & $M_{\rm tot}/L_{\rm bol}^{0.6}$ & \\ 
Source & [M$_{\sun}$] & [L$_{\sun}$] & [K] & [K] & $M_{\rm cold}/M_{\rm tot}$ & $L_{\rm cold}/L_{\rm bol}$ & $L_{\rm submm}/L_{\rm bol}$ & [M$_{\sun}$/L$_{\sun}^{0.6}$] & Class\\
\hline
IRAS 05399-0121 & $2.8\pm0.3$ & $21\pm1.2$ & $18.5\pm0.1$ & $103.9\pm0.2$ & $\sim1$ & 0.90 & 0.02 & 0.45 & 0/I\\
SMM 3 & $7.2\pm2.2$ & $3.5\pm0.2$ & $11.8\pm0.9$ & $36.9\pm0.2$ & $\sim1$ & 0.74 & 0.11 & 5.57 & 0\\
IRAS 05405-0117 & $1.6\pm0.2$ & $6.4\pm0.4$ & $16.1\pm0.1$ & $112.4\pm0.4$ & $\sim1$ & 0.69 & 0.03 & 0.53 & 0\\
SMM 4 & $3.8\pm0.2$ & $1.7\pm0.2$ & $11.6\pm0.2$ & $51.4\pm4.9$ & $\sim1$ & 0.76 & 0.11 & 2.76 & 0\\
IRAS 05412-0105 & $1.3\pm0.3$ & $5.8\pm0.6$ & $17.0\pm0.3$ & $127.8\pm0.5$ & $\sim1$ & 0.86 & 0.03 & 0.45 & 0\\
IRAS 05413-0104 & $1.0\pm0.2$ & $13.2\pm1.2$ & $20.3\pm0.4$ & $152.3\pm1.1$ & $\sim1$ & 0.97 & 0.02 & 0.21 & 0\\
\hline 
\end{tabular} 
\end{minipage}
\end{table*}

\subsection{Linear sizes, mass estimates and densities}

The linear sizes (radii $R=\theta_{\rm s}d/2$) were computed from the
angular FWHM sizes listed in Table~\ref{table:cores}.

The masses of the cores (gas+dust mass, $M_{\rm cont}$) are calculated
from their integrated 870 $\mu$m continuum flux density, $S_{870}$, assuming
that the thermal dust emission is optically thin:

\begin{equation}
\label{eq:mass}
M_{\rm cont}=\frac{S_{870}d^2}{B_{870}(T_{\rm d})\kappa_{870}R_{\rm d}} \; ,
\end{equation}
where $d$ is the distance, and $B_{870}(T_{\rm d})$ is the Planck function
with dust temperature $T_{\rm d}$.
For all IRAS sources as well as for SMM 3 and SMM 4 we adopted the dust
temperatures resulting from the SED fitting (see Table~\ref{table:sed},
column (4)). 
For all the other sources it was assumed that $T_{\rm d}=10$ K. 
The assumed dust temperature of 10 K is justified by
the estimates obtained from NH$_3$ (\cite{HWW1993}) and is commonly adopted 
for starless cores. The assumption that $T_{\rm d}=T_{\rm kin}$, where 
$T_{\rm kin}$ is the gas kinetic temperature, is likely to
be valid at densities $n({\rm H_2})>10^5$ cm$^{-3}$ (\cite{burke1983}).
The opacity per unit mass column density at
$\lambda=870$ $\mu{\rm m}$ is assumed to be
$\kappa_{870}\simeq0.17$ m$^2$ kg$^{-1}$. This value is interpolated from 
Ossenkopf \& Henning (1994, see Sect. 4.1). The value $1/100$ is adopted
for the dust-to-gas mass ratio, $R_{\rm d}$.

The virial masses, $M_{\rm vir}$, of IRAS 05405-0117 and Ori B9 N have been
estimated by approximating the mass distribution by a homogenous, isothermal
sphere without magnetic support and external pressure (see, e.g., Eqs.~(1) and
(2) in Chen et al. (2008) where the linewidth of N$_2$H$^+$ is used).
The resulting virial masses are about 4.3
M$_{\sun}$ for IRAS 05405-0117 and 2.8 M$_{\sun}$ for Ori B9 N.
The corresponding $M_{\rm cont}/M_{\rm vir}$ ratios are about 0.3 and 1.4. 
Note that it is usual that protostellar cores, like IRAS 05405-0117, appears 
to be below the self-gravitating limit ($M_{\rm cont}/M_{\rm vir}=0.5$),
though they are forming stars (e.g., \cite{enoch2008}).
Since Ori B9 N appears to be
gravitationally bound, it is probably prestellar.
There are several factors that would lead to virial masses being
overestimated. For example, using the radial density profile with power-law 
indices $p=1-1.5$ (see Sect. 5.3), would lead to $M_{\rm vir}$ being reduced
by factors of $1.1-1.25$.  

The volume-averaged H$_2$ number densities, $\langle n({\rm H_2}) \rangle$,
were calculated assuming a spherical geometry for the sources and using masses,
$M_{\rm cont}$, and radii, $R$, estimated for them from the dust continuum map.
The obtained radii, masses, and volume-averaged H$_2$ number densities are
given in colums (2), (3), and (5) of Table~\ref{table:parameters},
respectively. 

\begin{table*}
\caption{Linear radii, masses, and H$_2$ column and volume-averaged number
densities of all detected submm sources.}
\begin{minipage}{2\columnwidth}
\centering
\renewcommand{\footnoterule}{}
\label{table:parameters}
\begin{tabular}{c c c c c}
\hline\hline 
       & $R$ & $M_{\rm cont}$ & $N({\rm H_2})$ & $ \langle n({\rm H_2}) \rangle $\\ 
Source & [pc] & [M$_{\sun}$] & [$10^{22}$ cm$^{-2}$] & [$10^5$ cm$^{-3}$]\\  
\hline
IRAS 05399-0121 & 0.03 & 3.6 & 2.61 & 7.5\\ 
SMM 1    & 0.06 & 11.8 & 3.89 & 3.1\\
SMM 2    & 0.03 & 2.8 & 1.99 & 5.8\\
SMM 3    & 0.02 & 7.2 & 7.80 & 51.1\\
IRAS 05405-0117 & 0.05\footnote{These values include both the IRAS 05405-0117 and Ori B9 E.} & 1.5$^a$ & 0.76 & 0.7$^a$\\
SMM 4   & 0.04 & 3.8 & 1.98 & 3.4\\
SMM 5   & 0.04 & 2.3 & 1.51 & 2.1\\
SMM 6    & 0.10 & 9.9 & 2.45 & 0.6\\
Ori B9 N        & 0.05 & 3.9 & 1.51 & 1.7\\
SMM 7   & 0.03 & 3.1 & 3.01 & 6.5\\
IRAS 05412-0105 & -\footnote{Deconvolving the angular size was not possible, and thus the radius and number density could not be estimated.} & 0.8 & 0.63 & -$^b$\\
IRAS 05413-0104 & 0.03 & 1.0 & 1.85 & 2.2\\
\hline 
\end{tabular} 
\end{minipage}
\end{table*}

\subsubsection{Total mass of the Ori B9 region}

We made an estimate of the total mass in the region by using the 
near-infrared extinction mapping technique
(NICER, \cite{lombardi2001}).
In this technique, the near-infrared colors, namely $H-K$ and
$J-H$, of the stars shining through the dust cloud are compared to the colours
of stars in a nearby field that is free from dust.
The reddened colours of the stars behind the dust cloud can then be interpreted
in terms of extinction due to the relatively well-known ratios of optical
depths at $JHK$ wavelengths (for further details of the method, we refer to
Lombardi \& Alves (2001)). To implement the method, we retrieved
$JHK$ photometric data from the 2MASS archive, covering a
$30\arcmin \times 19\arcmin$
region centred at $(\alpha, \delta)_\mathrm{J2000}$ = (5:43:00, -01:16:20).
Applying NICER to these data yielded an extinction map with the
resolution of FWHM=$2\farcm5$, indicating extinction values of
$A_\mathrm{V}=8\dots 12$ mag at the positions of the detected sources.

The total mass of the region was calculated from the derived extinction map by
summing up the extinction values of all pixels assuming the gas-to-dust ratio
of $N(\mathrm{H}) = 2\cdot 10^{21}$ cm$^{-2}$ mag$^{-1}$
(\cite{bohlin1978}), and the mean molecular weight per H$_2$ molecule of 2.8. 
The total mass of the region resulting from the calculation is 1400 M$_{\sun}$.
The total mass of the cores within the region implied by the submm dust 
emission data is only $\sim50$ M$_{\sun}$, about $3.6$\% of the total mass in
the region. 

\subsection{Column densities, fractional abundances, and the degree of 
deuterium fractionation}

The H$_2$ column densities, $N({\rm H_2})$, towards the submm peaks and the
positions selected for the line observations were calculated using the
following equation:

\begin{equation}
\label{eq:N_H2}
N({\rm H_2})=\frac{I_{870}^{\rm dust}}{B_{870}(T_{\rm d})\mu_{\rm H_2} m_{\rm H}\kappa_{870}R_{\rm d}} \, .
\end{equation}
$I_{870}^{\rm dust}$ is the observed dust peak surface brightness, which 
is related to the peak flux density via 1 Jy/18\farcs6 beam
$=1.085\cdot10^{-18}$ W m$^{-2}$ Hz$^{-1}$ sr$^{-1}$.
$\mu_{\rm H_2}=2.8$ is the mean molecular weight per H$_2$ molecule, and
$m_{\rm H}$ is the mass of the hydrogen atom.
The same dust temperature values were used as in the mass estimates 
(Eq.~(\ref{eq:mass})).

The N$_2$H$^+$ column densities were calculated using the 
equation

\begin{equation}
\label{eq:N_tot}
N_{\rm tot}=\frac{3\epsilon_0 h}{2\pi^2 \mu_{\rm el}^2}\frac{1}{S_{\rm ul}}e^{E_{\rm u}/k_{\rm B}T_{\rm ex}}F(T_{\rm ex})Z(T_{\rm ex})\int \tau ({\rm v}) {\rm dv} \, ,
\end{equation}
where $\epsilon_0$ is the vacuum permittivity, $\mu_{\rm el}$ is the permanent
electric dipole moment, $S_{\rm ul}$ is the line strength, $E_{\rm u}$ is the
upper state energy, $Z$ is the rotational partition function, and 
$\int \tau {\rm dv}$ is the integrated optical thickness. 
We assumed a dipole moment of 3.4 D for both N$_2$H$^+$ and N$_2$D$^+$
(\cite{havenith1990}). For the rotational transition
$J_{\rm u} \rightarrow J_{\rm u}-1$ of a linear molecule (like N$_2$H$^+$),
$S_{\rm ul}=J_{\rm u}$.

For the N$_2$H$^+$ lines the optical thicknesses
were derived from the Gaussian fits to the hyperfine components, and thus
the integral $\int \tau {\rm dv}$ can be replaced by 
$\frac{\sqrt{\pi}}{2\sqrt{\ln 2}}\Delta {\rm v}\tau_0$. Here
$\Delta {\rm v}$ is the linewidth of an individual hyperfine component, and
$\tau_0$ is the sum of the peak optical thicknesses of all the seven
components. 

The N$_2$D$^+$ column densities were calculated in two different ways: 1) as
in the case of N$_2$H$^+$, and 2) using Eq.~(1) with the approximation
of optically thin line ($\tau \ll 1$):

\begin{equation}
\label{eq:ant_2}
T_{\rm A}^{*}\approx \eta \frac{h\nu}{k_{\rm B}}\left[F(T_{\rm ex})-F(T_{\rm bg})\right]\tau \,.
\end{equation}
Again, we assumed that $\eta=\eta_{\rm MB}$.
The integrated opacity was estimated from the integrated $T_{\rm MB}$ of the
main hyperfine group (54.3\% of the total integrated intensity):

\begin{equation}
\label{eq:tau}
\int \tau {\rm dv}=\frac{\int T_{\rm MB}{\rm dv}}{\frac{h\nu}{k_{\rm B}}\left[F(T_{\rm ex})-F(T_{\rm bg})\right]} \,.
\end{equation}

A comparison of the column density determination
via the two methods shows that $N({\rm N_2D^+})$, when using the first method,
is 1.3 (IRAS 05405-0117), 0.7 (Ori B9 E), and 0.6 (Ori B9 N) times the value
obtained using the second one.
As the N$_2$D$^+$ line areas are somewhat uncertain,
the N$_2$D$^+$ column densities determined by using the first method have been
adopted in this paper.  

The fractional N$_2$H$^+$ and N$_2$D$^+$ abundances, $x({\rm N_2H^+})$ and
$x({\rm N_2D^+})$, were calculated by dividing the corresponding column
densities by $N({\rm H_2})$ from the dust continuum. For $x({\rm N_2H^+})$ 
the dust map was smoothed to 26\farcs4, the resolution of the N$_2$H$^+$
observations. No smoothing was done in the case of N$_2$D$^+$, as the
resolutions of the N$_2$D$^+$ and dust continuum observations are
similar (16\farcs0 and 18\farcs6, respectively).
The degree of deuterium fractionation in N$_2$H$^+$ is defined as the
column density ratio $R_{\rm deut} \equiv N({\rm N_2D^+})/N({\rm N_2H^+})$.

The obtained H$_2$ column densities are given in column (4) of 
Table~\ref{table:parameters}. The N$_2$H$^+$ and N$_2$D$^+$
column densities, fractional abundances, and the values of
$R_{\rm deut}$ are listed in Table~\ref{table:d_frac}.
The uncertainties on $N({\rm N_2H^+})$ and
$N({\rm N_2D^+})$ have been calculated by propagating the uncertainties on
$T_{\rm ex}$, $\tau_{\rm tot}$, and $\Delta {\rm v}$, and the uncertainties on 
$N({\rm N_2D^+})/N({\rm N_2H^+})$ ratios are propagated from
$N({\rm N_2H^+})$ and $N({\rm N_2D^+})$.

\begin{table*}
\caption{N$_2$H$^+$ and N$_2$D$^+$ column densities, fractional abundances,
and the column density ratio.}
\begin{minipage}{2\columnwidth}
\centering
\renewcommand{\footnoterule}{}
\label{table:d_frac}
\begin{tabular}{c c c c c c}
\hline\hline 
         & $N({\rm N_2H^+})$ & $N({\rm N_2D^+})$ & $x({\rm N_2H^+})$ & $x({\rm N_2D^+})$ & \\
Position & [$10^{12}$ cm$^{-2}$] & [$10^{11}$ cm$^{-2}$] & [$10^{-10}$] & [$10^{-11}$] & $R_{\rm deut} \equiv N({\rm N_2D^+})/N({\rm N_2H^+})$\\
\hline
IRAS 05405-0117 & $9.14\pm0.08$\footnote{Harju et al. (2006) estimated 
slightly lower N$_2$H$^+$ column density, $\sim6-8\cdot10^{12}$ cm$^{-2}$, toward IRAS 05405-0117 from
the N$_2$H$^+(1-0)$ data of Caselli \& Myers (1994).} & $3.19\pm0.37$ & 11.1 & 4.9 & $0.03\pm0.004$\\
Ori B9 E & $4.54\pm0.65$ & $1.90\pm1.39$ & 6.9 & 5.0 & $0.04\pm0.03$\\
Ori B9 N\footnote{For the other velocity component $N({\rm N_2H^+})=1.56\pm0.09\cdot10^{12}$ cm$^{-2}$, $N({\rm N_2D^+})=7.64\pm5.44\cdot10^{10}$ cm$^{-2}$, and $R_{\rm deut}=0.05\pm0.03$.} & $4.11\pm1.51$ & $1.46\pm0.65$ & 3.9 & 1.9 & $0.04\pm0.02$\\
\hline 
\end{tabular} 
\end{minipage}
\end{table*}

\begin{table*}
\caption{Parameters derived in Sect. 4.4.}
\begin{minipage}{2\columnwidth}
\centering
\renewcommand{\footnoterule}{}
\label{table:ion}
\begin{tabular}{c c c c c c c c}
\hline\hline 
Source & $x({\rm H_2D^+})$ & $x({\rm H_3^+})$ & $x({\rm e})_l$\footnote{The first value is calculated from Eq.~(\ref{eq:low_x(e)}), whereas the second value is calculated from Eq.~(\ref{eq:lowxe2}).} & $x({\rm e})_u$ & $\langle x({\rm e}) \rangle$\footnote{This is the mean value between the lower and upper limit, where $x({\rm e})_l$ is calculated from Eq.~(\ref{eq:lowxe2}).} & $\zeta_{\rm H_2}$\footnote{The second value is derived by including HCO$^+$ in the analysis (see text).} & $C_i$\\
 & [$10^{-9}$] & [$10^{-8}$] & [$10^{-8}$] & [$10^{-7}$] & [$10^{-7}$] & [$10^{-16}$ s$^{-1}$] & [$10^3$ cm$^{-3/2}$ s$^{1/2}$]\\
\hline
Ori B9 E & 2.0 & 1.8 & 2.0/3.0 & 6.4 & 3.4 & 2.0/1.0 & 2.6\\
Ori B9 N & 1.0 & 0.9 & 1.0/2.0 & 6.3 & 3.3 & 2.5/1.3 & 4.1\\
\hline
\end{tabular} 
\end{minipage}
\end{table*}

\subsection{Ionization degree and cosmic ray ionization rate}

The charge quasi-neutrality of plasma dictates that the number of
positive and negative charges are equal.  Since electrons are the
dominant negative species, their fractional abundance nearly equals
the sum of the abundances of positive ions, $x({\rm cations})\simeq
x({\rm e})$. Thus, one may obtain a \textit{lower limit} for the
ionization fraction by summing the abundances of several molecular
ions:

\begin{equation}
\label{eq:low_x(e)}
x(\mathrm{e}) >
x(\mathrm{N_2H^+})+x(\mathrm{N_2D^+})+x(\mathrm{H_3^+})+x(\mathrm{H_2D^+}) \; .
\end{equation}

In the following we attemp to derive estimates for the cosmic ray
ionization rate and the fractional electron abundance using the
abundances of $\mathrm{N_2H^+}$, $\mathrm{N_2D^+}$, and
$\mathrm{H_2D^+}$ together with the reaction schemes and formulae
presented in Crapsi et al. (2004) and Caselli et al. (2008). The rate
coefficients for the ${\rm H_3^+ + H_2}$ isotopic system have been
newly calculated by Hugo et al. (2009). We use these for the
deuteration sequence ${\rm H_3^+}$ $\leftrightarrow$ ${\rm H_2D^+}$
$\leftrightarrow$ ${\rm D_2H^+}$ $\leftrightarrow$ ${\rm D_3^+}$ (see
their Table VIII) . For other reactions the rate coefficients have
been adopted from the UMIST database which is available at {\tt www.udfa.net}. 
The main difference between the Hugo et al. coefficients and those of
Roberts et al. (2004) is that in the former, the effective backward
rate coefficient, $k_{-1}$, of the reaction ${\rm H_3^+}+{\rm HD}
\Harpoons^{k_1}_{k_{-1}}{\rm H_2D^+}+{\rm H_2}$, and the corresponding
coefficients for multiply deuterated forms of H$_3^+$ are higher if
the non-thermal ortho/para ratio of H$_2$ (hereafter o/p-H$_2$) is
taken into account (\cite{pagani1992}; \cite{gerlich2002}; \cite{flower2006a}; 
\cite{pagani2009a}; \cite{hugo2009}).

The ortho-H$_2$D$^+$ column density was derived towards Ori B9 E and N
by Harju et al. (2006). Using their value, $N({\rm o-H_2D^+}) \sim
3.0\cdot10^{12}$ cm$^{-2}$, and the H$_2$ column densities derived
here, we get $x({\rm o-H_2D^+})\approx8.0\cdot10^{-10}$ and
$4.1\cdot10^{-10}$ towards Ori B9 E and N,
respectively\footnote{Caselli et al. (2008) derived $N({\rm
o-H_2D^+})= 2.0/9.0\cdot10^{12}$ cm$^{-2}$ toward position which is
only 12\farcs7 southeast of our line observations position Ori B9 N,
assuming a critical density $n_{\rm cr}=10^5$ and $10^6$ cm$^{-3}$,
respectively.}.  The ortho/para ratio of H$_2$D$^+$ (hereafter
o/p-${\rm H_2D^+}$) depends heavily on o/p-H$_2$. According to the
model of Walmsley et al. (2004, see their Fig.~3), the characteristic
steady-state value of o/p-H$_2$ is $\sim 10^{-4}$ in the density range
$n({\rm H_2}) \sim 10^5 - 10^6$ cm$^{-3}$ appropriate for the objects
of this study. This model deals with the situation of ``complete
depletion'' and it is not clear how valid the quoted o/p-H$_2$ is in
less depleted gas. The recent results of Pagani et al. (2009a) suggest
high values of o/p-H$_2$ ($\sim 4\cdot10^{-3} - 6\cdot10^{-2}$) in
L183.

For the moment we adopt the value o/p-${\rm H_2}=10^{-4}$. The
effect of increasing this ratio will be examined briefly
below. Assuming that o/p-${\rm H_2D^+}$ is mainly determined by
nuclear spin changing collisions with ortho- and para-H$_2$, the
quoted o/p-${\rm H_2}$ ratio implies an o/p-${\rm H_2D^+}$ of $\sim
0.7$ at $T=10$ K. The total (ortho$+$para) H$_2$D$^+$ abundances
corresponding to this o/p ratio are $x({\rm
H_2D^+})\approx2.0\cdot10^{-9}$ and $1.0\cdot10^{-9}$ towards Ori B9 E
and N, respectively.

The N$_2$D$^+$/N$_2$H$^+$ column density ratio which we denote by
$R_{\rm deut}$, gives a rough estimate for the H$_2$D$^+$/H$_3^+$
abundance ratio, denoted here by $r$. According to the relation 
$R_{\rm deut}\approx(r+2r^2)/(3+2r+r^2)$ derived by Crapsi et al. (2004; a
more accurate formula is given in Eq.~(13) of Caselli et al. (2008)),
the values of $R_{\rm deut}$ given in Table~\ref{table:d_frac} imply
$r\approx0.08$ for IRAS 05405-0117, and $r\approx0.11$ for Ori B9 E and
N.  Using these $r$ values we obtain the following fractional
abundances $x({\rm H_3^+})=x({\rm H_2D^+})/r$ $\approx1.8\cdot10^{-8}$
(Ori B9 E) and $\approx9.1\cdot10^{-9}$ (Ori B9 N). Substituting all
the derived abundances into Eq.~(\ref{eq:low_x(e)}) we get the
following lower limits for the degree of ionization: $x({\rm
e})>2.0\cdot10^{-8}$ in Ori B9 E, and $>1.0\cdot10^{-8}$ in Ori B9 N.

Despite the fact that the clump associated with IRAS 05405-0117 does
not stand out in the $^{13}$CO and C$^{18}$O maps, a moderate CO
depletion factor of 3.6 near Ori B9 N has been derived (\cite{caselli1995}; 
\cite{caselli2008}). This estimate is based on
$^{13}{\rm CO}(1-0)$ observations made with the FCRAO 14-m telescope (HPBW
$50\arcsec$), and a total H$_2$ column density, $N({\rm H_2})$,
derived from ammonia. By smoothing the LABOCA map to the resolution of
$50\arcsec$, we obtain an average H$_2$ column density of
$1.3\cdot 10^{22}$ cm$^{-2}$ around Ori B9 N. This is 2.6 times lower
than the value adopted by Caselli et al. (2008). 
With this $N({\rm H_2})$ the fractional CO abundance, $x({\rm CO})$, 
becomes $6.8\cdot10^{-5}$. The corresponding CO depletion factor, $f_{\rm D}$,
is only 1.4 with respect to the often adopted fractional abundance from 
Frerking et al. (1982)\footnote{The depletion factor is used in the 
text to express the fractional CO abundance with respect to the value 
$9.5\cdot10^{-5}$. Adopting a higher reference abundance (see \cite{lacy1994})
would not change the results of the calculations.}.
The small value of $R_{\rm deut}$ is consistent with a low CO depletion 
factor (e.g., \cite{crapsi2004}). 

In chemical equilibrium the fractional ${\mathrm H_3^+}$ abundance is 

\begin{equation}
x(\mathrm{H_3^+})=\frac{\zeta_{\mathrm{H_2}}/n(\mathrm{H_2})+
k_{-1} x(\mathrm{H_2D^+})}{D_0} \; , 
\label{eq:h3+}
\end{equation} 
where  $\zeta_{\rm H_2}$ is the cosmic ray ionization rate of H$_2$,
and

\begin{displaymath}
D_0 \equiv  k_1 x(\mathrm{HD}) + k_{\rm CO} x({\rm CO}) +k_{\rm rec0} x({\rm e}) + k_{\rm g} x({\rm g^-}) +k_{\rm N_2} x({\rm N_2}) + ...
\end{displaymath}
The notation of Caselli et al. (2008) has been used here, i.e., $k_1$
and $k_{-1}$ are the forward and backward rate coefficients of the
reaction mentioned above, and the other terms in $D_0$ refer to the
destruction of H$_3^+$ in reactions with neutral molecules (e.g., CO
and N$_2$) and in recombination with electrons and on negatively
charged dust grains.

By solving numerically Eqs.~(8)-(10), and (13) of Caselli et
al. (2008) together with our Eq.~(\ref{eq:h3+}) we obtain the following
estimates for the fractional electron abundance and cosmic ray ionizations
rate: $x({\rm e}) = 6.4\cdot 10^{-7}$, $\zeta_{\rm H_2} = 2.0\cdot10^{-16}$
s$^{-1}$ in Ori B9 E, and $x({\rm e}) = 6.3 \cdot 10^{-7}$, $\zeta_{\rm H_2} =
2.5\cdot10^{-16}$ s$^{-1}$ in Ori B9 N. Here we have used CO
depletion factor 1.4, and the dust parameters ($k_{\rm g^-}$, $x({\rm
g^-}$) quoted in Eqs.~(11) and (12) of Caselli et al. which are based
on a MRN dust grain size distribution (\cite{mathis1977}) and
effective grain recombination coefficients derived by Draine \& Sutin
(1987). The average number densities, $\langle n({\rm H_2}) \rangle$,
derived in Sect. 4.2. have been used for the cosmic ray ionization rates.
The obtained values of $\zeta_{\rm H_2}$ are very similar to each other as
is expected for such a nearby cores (\cite{williams1998};
\cite{bergin1999} and references therein).

For comparison, Bergin et al. (1999) found that adopting $\zeta_{\rm
H_2}=5\cdot10^{-17}$ s$^{-1}$ in their chemical model best reproduced
their observations of massive cores in Orion.  Note that the
``standard'' value often quoted in the literature is 
$\zeta_{\rm H_2}=1.3\cdot10^{-17}$ s$^{-1}$.
Also the fractional electron abundances are clearly larger than those
calculated from the standard relation $x({\rm
e})\sim1.3\cdot10^{-5}n({\rm H_2})^{-1/2}$ (cf. \cite{mckee1989};
\cite{mckee1993}), where the electron fraction is due to cosmic ray ionization
only and $\zeta_{\rm H_2}$ has its above mentioned standard value. 
The corresponding values would be $x({\rm e})\sim 5\cdot10^{-8}$ 
(Ori B9 E) and $x({\rm e}) \sim 3\cdot10^{-8}$ (Ori B9 N).
The mean value of the ionization degree found by Bergin et
al. (1999) for the massive cores in Orion is $\sim 8 \cdot 10^{-8}$.
 
The parameters derived above depend on the adopted o/p-H$_2$ which
affects the backward rate coefficient $k_{-1}$, $k_{-2}$, and $k_{-3}$ 
(see \cite{caselli2008}), and the CO depletion factor $f_{\rm D}$ which
affects the destruction of H$_3^+$. The fractional electron
abundance can be decreased to $\sim 8\cdot10^{-8}$ by increasing
o/p-H$_2$ to $2.4\cdot10^{-3}$ (this yields a $\zeta_{\rm H_2}$ of 
$1.3\cdot10^{-16}$ s$^{-1}$). On the other hand, an increase in
$f_{\rm D}$ will lead to a higher $x({\rm e})$, but also to a lower
$\zeta_{\rm H_2}$. A solution where both $x({\rm e})$ and
$\zeta_{\rm H_2}$ obtain the average values derived by Bergin et al.
(1999) can be found by setting $f_{\rm D}$ to 4.4 and o/p-H$_2$ to
$3.4\cdot10^{-3}$.

However, the available observational data do not give grounds for
abandoning the present bona fide $f_{\rm D}$ value 1.4. So the
main uncertainty seems to be related to the unknown o/p-H$_2$.
Additional uncertainties to the $\zeta_{\rm H_2}$ values are caused by the
rough density estimates, and by the fact that densities in the positions
observed in molecular lines are probably lower than the average
densities adopted in the analysis. The electron abundance obtained assuming an
o/p-${\rm H_2}$ of $1.0\cdot10^{-4}$ is likely to be an upper limit. This o/p
ratio corresponds to steady state in highly depleted dense gas with
large abundances of H$^+$ and H$_3^+$ capable of efficient proton
exchange with H$_2$ (\cite{flower2007}). Their replacement by other
ions in less extreme situations can sustain higher o/p-H$_2$ ratios. 

A substantial amount of CO implies the presence of HCO$^+$ in the gas.
By including the dissociative electron recombination of HCO$^+$, and
the proton exchange reaction between N$_2$H$^+$ (or N$_2$D$^+$) and CO
in the reaction scheme, the fractional HCO$^+$ abundance can be
solved. Through this estimate we get a slightly more stringent lower
limit on the electron abundance than that imposed by
Eq.~(\ref{eq:low_x(e)}) by demanding that

\begin{align}
x({\rm e}) & \geq  x({\rm H_3^+})(1+r) + x({\rm N_2H^+})(1+R_{\rm deut})  \nonumber \\ 
           &\quad+ x({\rm HCO^+})(1+R_{\rm deut}) \; .
\label{eq:lowxe2}
\end{align}
Here it has been assumed that N$_2$H$^+$ and HCO$^+$ have
similar degrees of deuterium fractionation. By varying o/p-H$_2$
until electron and the ``known'' cations are in balance we obtain with
o/p-${\rm H_2} = 2.7\,10^{-3}$ the lower limits $x({\rm e}) \geq
3\cdot10^{-8}$ and $x({\rm e}) \geq 2\cdot10^{-8}$ in Ori B9 E and N,
respectively. The corresponding values of $\zeta_{\rm H_2}$ are
$1.0\cdot10^{-16}$ s$^{-1}$ and $1.3\cdot10^{-16}$ s$^{-1}$.  In both
solutions HCO$^+$ is more abundant than H$_3^+$, whereas in the case
of a large $x({\rm e})$ (small o/p-${\rm H_2}$) the HCO$^+$ abundance
lies between those of N$_2$H$^+$ and H$_2$D$^+$.

The obtained values of the cosmic ray ionization rate vary smoothly
with o/p-${\rm H_2}$, and all viable solutions point towards $\zeta_{\rm
H_2} \sim 1-2\cdot10^{-16}$ s$^{-1}$. In the model of McKee (1989) these
levels imply factional ionizations of $\sim 1.1-1.6 \cdot10^{-7}$ at the 
density $10^5$ cm$^{-3}$. These values lie between the lower and upper
limits derived above. In what follows we assume that 
$x({\rm e}) \sim 1\cdot10^{-7}$, keeping in mind that true electron abundance
is likely to be found within a factor of few from this value.   

There is also uncertainty about the most abundant ion. According to
our calculation the electron abundance is an order of magnitude higher than
the summed abundances of the positive ions
H$_3^+$, HCO$^+$, and N$_2$H$^+$ for ${\rm o/p-H}_2 =
1.0\cdot10^{-4}$, whereas for the higher o/p-H$_2$ ratio $x({\rm
HCO^+})$ is comparable to $x({\rm e})$. This suggests that in the
first case the reaction scheme misses the most abundant cation(s). In
depleted regions with densities below $10^6$ cm$^{-3}$ protons, H$^+$,
are likely to be the dominant ions (\cite{walmsley2004}; \cite{pagani2009a}).
On the other hand, as discussed in Crapsi et al. (2004)
and references therein, if atomic oxygen is abundant in the gas phase
the major ion may be H$_3$O$^+$. To our knowledge this ion has not yet been
found in cold clouds (see also Caselli et al. 2008).
When discussing the ambipolar diffusion timescale in Sect. 5.6 we will assume
that the most abundant ion is either H$^+$ or HCO$^+$.
 
We furthermore estimate the value of a constant, $C_i$, that
describes the relative contributions of molecular ions and metal ions
to the ionization balance (\cite{williams1998}, their Eq.~(4);
\cite{bergin1999};
\cite{padoan2004}). The value of $C_i$ can be used to
estimate the strength of the ion-neutral coupling in terms of the wave
coupling parameter, $W \propto C_i$ (see Sect. 5.7).
In this analysis it is assumed that the electron abundance is
determined by cosmic ray ionization balanced by recombination and
it is appropriate for cores where $A_{\rm V}>4$ mag (i.e., ionization due to
cosmic rays dominates that resulting from UV radiation).
Adopting the electron abundance $1\cdot10^{-7}$ and 
$\zeta_{\rm H_2}\sim10^{-16}$ s$^{-1}$ we find values of 
$C_i\simeq2.6-4.1\cdot10^3$ cm$^{-3/2}$ s$^{1/2}$ in our cores.
These are similar to the value found by Bergin et al. (1999)
for the massive cores in Orion ($3.6\cdot10^3$ cm$^{-3/2}$ s$^{1/2}$).
McKee (1989) derives $C_i=3.2\cdot10^3$
cm$^{-3/2}$ s$^{1/2}$ for an idealised model of cosmic ray ionization and 
Williams et al. (1998) obtained $C_i=2.0\cdot10^3$ cm$^{-3/2}$ s$^{1/2}$ 
for low-mass cores.

All the parameters derived in this Section are summarised in
Table~\ref{table:ion}.

\section{Discussion}

\subsection{Nature of submm sources in Ori B9}

By combining the submm LABOCA and far-infrared Spitzer data, we can distinguish
starless cores from protostellar cores. In addition to the four IRAS sources
in the region, two of the new submm sources, namely SMM
3 and SMM 4, are clearly associated with Spitzer point sources and are
protostellar. The remaining six submm cores are starless.

IRAS 05399-0121 was previously classified as a Class I protostar 
(\cite{bally2002} and references therein). However, taking into account the
rather low bolometric (18.5 K) and kinetic temperatures 
(13.7 K, \cite{HWW1993}), and high values of $L_{\rm submm}/L_{\rm bol}$
(2\%) and $M_{\rm tot}/L_{\rm bol}^{0.6}$ (0.45 M$_{\sun}$/L$_{\sun}^{0.6}$),
we suggest the source is in a transition phase from Class 0 to Class I 
(see \cite{bontemps1996}; \cite{froebrich2005}).
This source is associated with the highly collimated jet HH 92
(\cite{bally2002}).

The SED of IRAS 05413-0104 derived here is consistent with its previous
classification as a Class 0 object (e.g., \cite{cabrit2007} and references
therein). The source is associated with the highly symmetric jet HH 212
(\cite{lee2006}, 2007; \cite{codella2007}; \cite{smith2007}; 
\cite{cabrit2007}). IRAS 05412-0105 and IRAS 05405-0117 have very similar SEDs,
and they, too, are likely to represent the Class 0 stage. The weak line wings
in the N$_2$H$^+(1-0)$ hyperfine lines of IRAS 05405-0117
(see Fig.~\ref{figure:n2h+}, top) could indicate the presence of outflow from
an embedded protostellar object.

$L_{\rm submm}/L_{\rm bol}$ ratio for both SMM 3 and SMM 4 is 11\%.
This together with low values of $T_{\rm bol}$ make these new submm sources
Class 0 candidates (e.g., \cite{froebrich2005} and references therein) that
are deeply embedded in a massive, cold envelope.
On a bolometric luminosity vs. temperature diagram these objects lie on the
evolutionary track for a Class 0 source with initially massive envelope 
(see Fig.~12 in Myers et al. (1998)).

The starless cores SMM 1, 2, 5, 6, 7, and Ori B9 N, are 
likely to be prestellar as their densities are relatively high
($0.6-5.8\cdot10^5$ cm$^{-3}$; see also Sect. 4.2).
The 24 $\mu$m Spitzer source near SMM 5 is probably not associated
with this core. It lies rather far form the core centre and
it is not detected at 70 $\mu$m (Fig.~\ref{figure:spitzer}).

There is an equal number of prestellar and protostellar cores
in Ori B9. This situation is similar to that recently found by
Enoch et al. (2008) in Perseus, Serpens, and Ophiuchus,
and suggests that the lifetimes of prestellar and protostellar
cores are comparable. Evolutionary timescales will be further discussed
in Sect. 5.6. 

\subsection{Mass distribution and core separations}

\begin{figure}[!h]
\resizebox{\hsize}{!}{\includegraphics{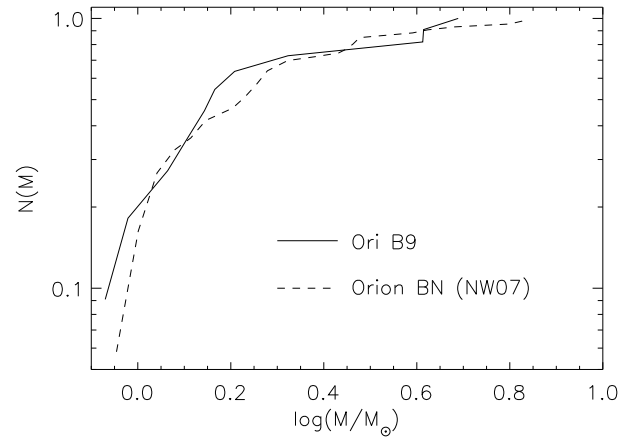}}
\caption{Normalised cumulative mass functions, $\mathcal{N}(M)$,
for the prestellar cores in Ori B9 (solid line) and in Orion B North 
(dashed line) studied by Nutter \& Ward-Thompson (2007).}
\label{figure:CMF}
\end{figure}

The spatial and mass distribution of cores are both important parameters
concerning the cloud fragmentation mechanism.
Our core sample is, however, so small that it is not reasonable to study 
the properties of these distributions directly. Therefore, we only compared
them with the distributions derived for another, larger core sample in Orion
GMC by NW07. We make this comparison particularly with Orion B North because
the SCUBA 850 $\mu$m map of Orion B North (see Fig.~2c in NW07) looks
qualitatively similar to Ori B9.
Orion B North also has deeper sensitivity and completeness limit than other 
regions studied by NW07, and besides it contains large number of cores.
Fig.~\ref{figure:CMF} presents the observed cumulative mass functions, which
counts cores with mass less than $M$, i.e., 
$\mathcal{N}(M)=N(m<M)/N_{\rm tot}$, for both the core masses in Ori B9 and
masses of prestellar cores in Orion B North derived by NW07.
Note that the core mass function (CMF) studied by NW07 is constructed by
removing the Class I protostars from the sample, so that CMF includes only 
cores which have all their mass initially available for star formation left.
Correspondingly, we have excluded IRAS 05399-0121 from our sample.
We have also multiplied the core masses by required factors to compare to NW07
values, due to differences in assumed values of $T_{\rm dust}$,
$\kappa_{\lambda}$, and distance (NW07 used the following values:
$T_{\rm dust}=20$ K, $\kappa_{850}=0.1$ m$^2$ kg$^{-1}$, and $d=400$ pc).

In order to determine if the two datasets are samples of the same core mass
distribution, we carried out the Kolmogorov-Smirnov (K-S) test.
The K-S test yields the maximum vertical difference between the cumulative
distributions of $D=0.166$, and probability of approximately 95\% that the
core mass distributions in Ori B9 and Orion B North are drawn from the same
parent distribution. 

\begin{figure}[!h]
\resizebox{\hsize}{!}{\includegraphics{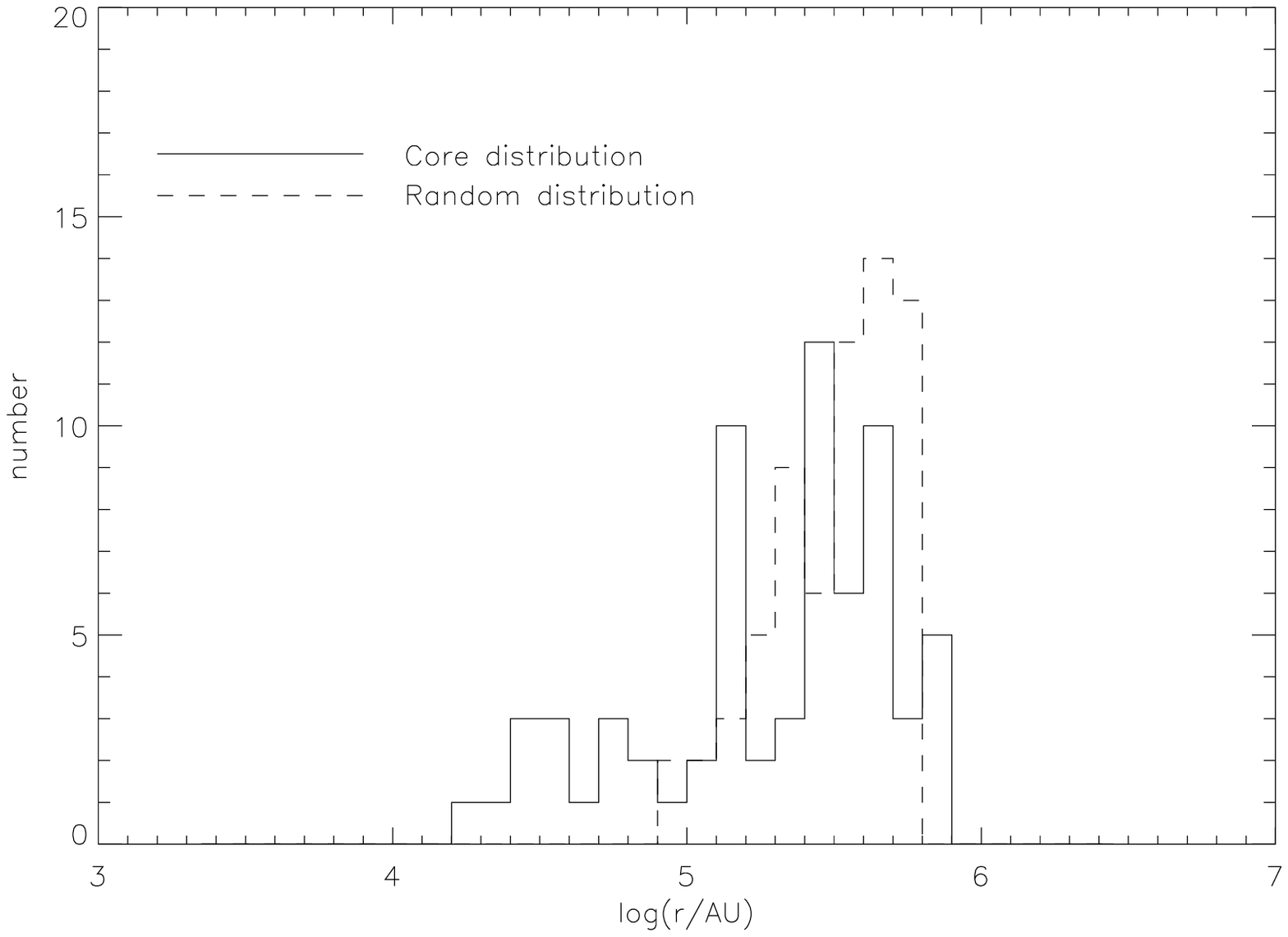}}
\resizebox{\hsize}{!}{\includegraphics{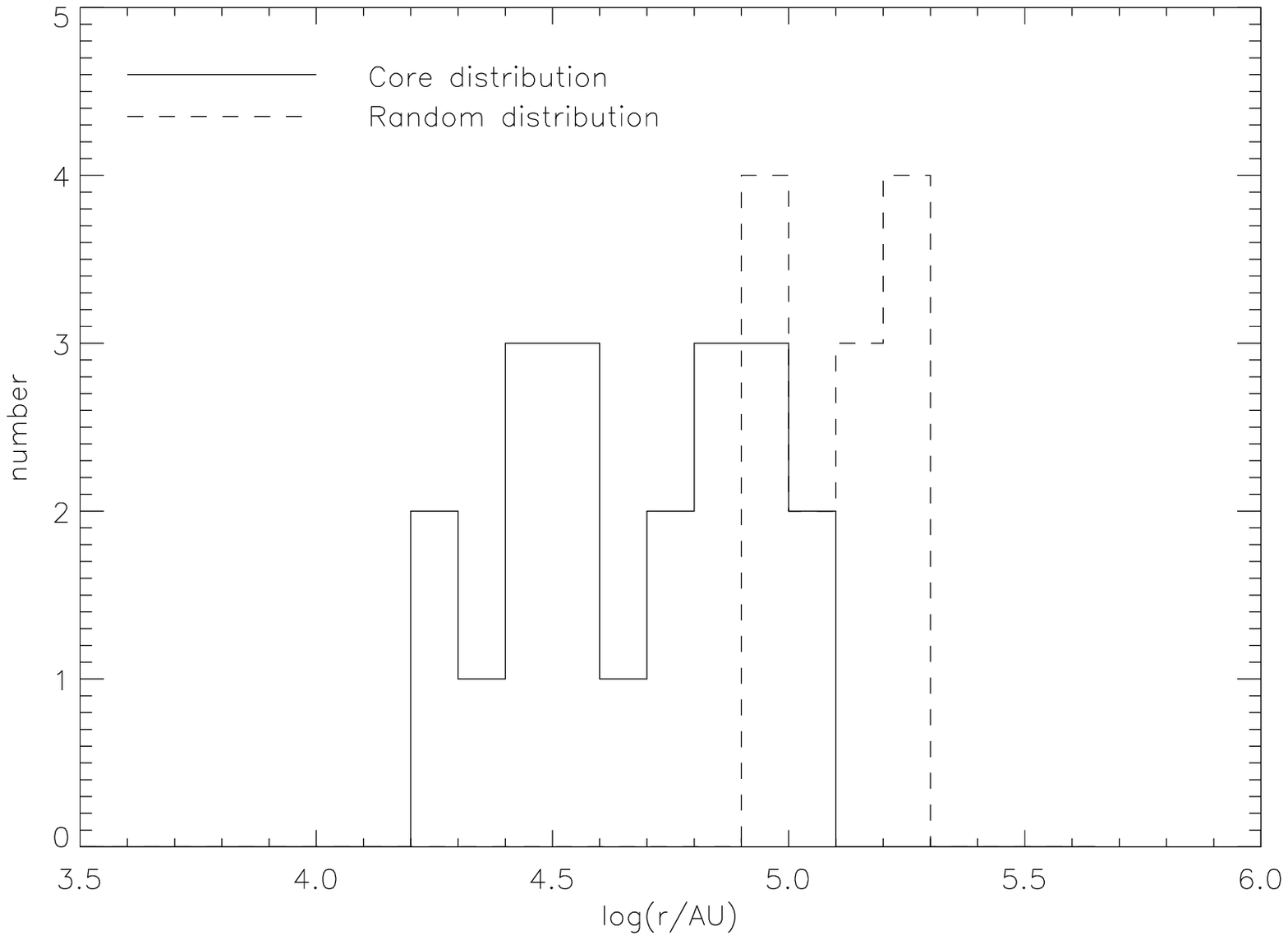}}
\caption{{\bf Top:} Observed core separation distribution (solid line) compared
with the expected distribution for random distribution of the same
number of sources as the observed sample over an identical area (dashed line).
{\bf Bottom:} Observed nearest neighbour distribution (solid line)
compared with the expected distribution for random distribution of the same
number of sources as the observed sample over the same area (dashed line).}
\label{figure:dist}
\end{figure}

Fig.~\ref{figure:dist} (top) shows the observed core
separation distribution and the distribution expected for the same number of
randomly positioned cores over an identical area (0.22 deg$^2$).
The mean and median of the core separations in Ori B9 
are $\log(r/{\rm AU})=5.47\pm0.09$ ($2.9\pm0.6\cdot10^5$ AU) and 5.42
($2.6\cdot10^5$ AU), respectively. The quoted error for the mean correspond 
to the standard deviation.
These values are similar to those of randomly positioned cores,
for which the mean and median are $\log(r/{\rm AU})=5.59\pm0.05$ and 
$5.59\pm0.06$, respectively. The quoted uncertainties are the standard
deviation of the sampling functions.
For core separation distribution in Orion B North studied by 
NW07, the corresponding values are $5.67\pm0.03$ and 5.63, suggesting that the
fragmentation scale is similar in both Ori B9 and Orion B North.
Similar fragmentation scales and the fact that CMFs have resemblance to the
stellar IMF (\cite{goodwin2008}) suggest that the origin of cores in these two
regions is probably determined by turbulent fragmentation 
(e.g., \cite{maclow2004}; \cite{ballesteros-paredes2007}).
The clustered mode of star formation in these two regions suggests that
turbulence is driven on large scales (e.g., \cite{klessen2001}).
Recently, Enoch et al. (2007) found the median separations
of $\log(r/{\rm AU})=3.79$, 4.41, and 4.36 in nearby molecular clouds
Ophiuchus, Perseus, and Serpens, respectively.
The spatial resolution of the Bolocam (31\arcsec) used by Enoch et al. at the
distance of Ophiuchus, Perseus, and Serpens, is 0.02, 0.04, and 0.04 pc.
The latter two are similar to our
resolution. The results suggest that the fragmentation scales in Perseus and
Serpens are different from that in Orion.

Fig.~\ref{figure:dist} (bottom) shows the comparison between the observed 
nearest neighbour distribution and the distribution for randomly positioned
cores. The mean and median of the nearest neighbour distribution in Ori B9 are 
$\log(r/{\rm AU})=4.75\pm0.09$ ($5.6\pm1.3\cdot10^4$ AU) and 4.62 
($4.2\cdot10^4$ AU), respectively. These values are rather different
from those expected from random distributions, for which the mean and median
are $\log(r/{\rm AU})=5.08\pm0.08$ and $5.07\pm0.11$, respectively.
For core positions in Orion B North the mean and median are
$\log(r/{\rm AU})=4.46\pm0.03$ and 4.35, respectively (NW07).
Also this comparison supports the idea that the scale of fragmentation,
and the amount of clustering are similar in Ori B9 and Orion B North. 
Note that the minimum observable separation is the beam size, i.e $18\farcs6$
or $\sim8.3\cdot10^3$ AU at 450 pc. We also note that the source sample is too
small to measure the significance of the clustering in Ori B9
based on the two-point correlation function. 

\subsection{Sizes, shapes, and density structures of the cores}

Starless cores in Ori B9, for which the mean value of the deconvolved angular 
size in units of the beam FWHM is
$\langle \theta_{\rm s}/\theta_{\rm beam} \rangle=2.5\pm0.8$, are larger on
average than protostellar cores
($\langle \theta_{\rm s}/\theta_{\rm beam} \rangle=1.6\pm0.2$).
These sizes are similar to those recently found by Enoch et al. (2008)
in Perseus ($\langle \theta_{\rm s}/\theta_{\rm beam} \rangle=2.2$ and 1.6 for 
starless and protostellar cores, respectively).
The mean axis ratios at half-maximum contours of starless and protostellar
cores are also 2.5 and 1.6, respectively (see Table~\ref{table:cores}, column
(7)). This indicates that starless cores in Ori B9 are also more
elongated on average than protostellar cores (cf. \cite{offner2009}).

The values of $\theta_{\rm s}/\theta_{\rm beam}$ can be used to infer the
steepness of the core radial density profile (\cite{young2003};
\cite{enoch2008}). According to the correlation between
$\theta_{\rm s}/\theta_{\rm beam}$ and density-power-law index, $p$, found
by Young et al. (2003, see their Fig.~27) a mean
$\theta_{\rm s}/\theta_{\rm beam}$ values of 2.5 for starless and 1.6 for
protostellar cores imply an average index of $p\sim0.9-1.0$ and
$\sim1.4-1.5$, respectively. 
Moreover, Fig.~25 of Young et al. (2003) suggest power-law indices $<1$ for
starless cores and $1.1-1.6$ for protostellar cores, consistent with those
inferred by the average deconvolved angular source sizes.
The low values of $p\sim1.0$ for starless cores suggest that they are 
best modelled with shallower density profiles than the protostellar cores.
These results are in agreement with those found by Ward-Thompson et al.
(1999) using 1.3 mm dust continuum data, and Caselli et al. (2002b) using
N$_2$H$^+(1-0)$ maps.

\subsection{Deuterium fractionation and depletion in
the IRAS 05405-0117 region}

The N$_2$D$^+$/N$_2$H$^+$ column density ratio, $R_{\rm deut}$, is supposed to
increase strongly as the core evolves (\cite{caselli2002}; \cite{crapsi2005a},
their Fig. 5; \cite{fontani2006}; \cite{emprechtinger2009};
but see \cite{roberts2007}). 
This can be understood so that the abundances of H$_3^+$, and its deuterated
forms which transfer deuterium to other molecules, increase with increasing
density due to molecular depletion and a lower degree of ionization. 
Crapsi et al. (2005a) suggested that prestellar cores are characterised with
$R_{\rm deut}>0.1$, whereas starless cores with $R_{\rm deut}<0.1$ are not
necessarily so dense that CO would be heavily depleted (\cite{roberts2007}).
It should be noted that in cores without internal heating sources
the degree of deuterium fractionation is likely to increase inwards
as the density increases and temperature decreases due to attenuation
of starlight. This temporal and radial tendency is likely to
be reversed during the core collapse because of compressional heating
and the formation of a protostar (e.g., \cite{aikawa2008a}; see also Fig.~3 
in Emprechtinger et al. (2009)).

The positions studied here have $R_{\rm deut}\sim0.03-0.04$. 
This is $\sim2-3\cdot10^3$ times larger than the cosmic
D/H elemental abundance of $\sim1.5\cdot10^{-5}$ (\cite{linsky1995}; 2006; 
\cite{oliveira2003}). Also, the H$_2$D$^+$/H$_3^+$ ratios we derived are
$\sim7\cdot10^3$ times larger than the cosmic D/H ratio.
Our $R_{\rm deut}$ values are similar to those found by Crapsi et al. (2005a)
toward several low-mass starless cores, and to those found by Emprechtinger et
al. (2009) toward Class 0 sources. 
Like Emprechtinger et al. (2009), we find that the deuterium fractionation of
N$_2$H$^+$ in protostellar cores, which takes place in the cold extended
envelope, is similar to that in prestellar cores.
It has been found that the values of $R_{\rm deut}$ toward high-mass
star-forming cores are (usually) lower than those found in the present
study (\cite{fontani2006}; see also \cite{emprechtinger2009}).
This conforms with the fact that our sources are low- to intermediate
mass star-forming cores.

As discussed by Walmsley et al. (2004) and Flower et al. (2006a),
the H$_2$D$^+$ abundance depends (inversely) on the
ortho:para ratio of H$_2$, because the reaction
${\rm H_2D^+}+{\rm H_2}\overset{k_{-1}}{\rightarrow}{\rm H_3^+}+{\rm HD}$
is rapid between ortho forms. The ortho:para ratio of H$_2$ decreases
with time and gas density, and is therefore large at early stages of core
evolution. Consequently, a relatively high degree of deuterium fractionation
is a sign of matured chemistry characterised by a low ortho:para ratio of
H$_2$ and probably a high degree of molecular depletion.
Low CO depletion factor of 1.4 close to N$_2$H$^+$
peak Ori B9 N (see Sect. 4.4) is consistent with the $R_{\rm deut}$ value of
Ori B9 N (e.g., \cite{crapsi2004}; see also Fig.~4 in Emprechtinger et al.
(2009)). 
The ortho-H$_2$D$^+$ detection towards Ori B9 E and N suggests an evolved
chemical stage and tells of a longlasting prestellar phase.
The non-detection toward IRAS 05405-0117 can be explained by a 
lower ortho-H$_2$D$^+$ abundance due to the central heating by the protostar.

\subsection{Evidence for a N$_2$H$^+$ ``hole'' and chemical differentiation}

The N$_2$H$^+$ map of Caselli \& Myers (1994, see their Fig.~2) and submm dust
continuum map of the clump associated with IRAS 05405-0117
(Fig.~\ref{figure:positions}) are not very much alike.
The strongest dust continuum peak, SMM 4, does not stand out in N$_2$H$^+$.
Moreover, the northern N$_2$H$^+$ maximum, Ori B9 N, seem to be shifted with
respect to the northern dust peak, and the N$_2$H$^+$ peak Ori B9 E does not
correspond to any dust emission peak.

To determine whether the $\sim1.5$ times higher resolution of the LABOCA 870 
$\mu$m map relative to the N$_2$H$^+$ map of Caselli \& Myers (1994) 
contributed to the different appearance of the dust continuum and N$_2$H$^+$
maxima, we smoothed the LABOCA map to a resolution similar to that of the
N$_2$H$^+$ map (27\arcsec). The smoothed 870 $\mu$m map, however, still shows
the same differences between the dust continuum and N$_2$H$^+$.

The Class 0 candidate SMM 4 (see Sect. 4.1 and 5.1) can represent an extreme
case of depletion where also N$_2$H$^+$ has disappeared from the gas phase due
to freeze out on to the dust grain surfaces. There is some previous evidence
for N$_2$H$^+$ depletion in the centres of chemically evolved cores,
such as B68 (\cite{bergin2002}), L1544 (\cite{caselli2002a}), L1512
(\cite{lee2003}), and L1521F (\cite{crapsi2004}).
Example of the N$_2$H$^+$ depletion toward 
Class 0 source is IRAM 04191+1522 in Taurus (\cite{belloche2004}).
Pagani et al. (2005) found clear signs of moderate N$_2$H$^+$ depletion in the
prestellar core L183 (see also \cite{pagani2007}).
Also, Schnee et al. (2007) found clear evidence
of N$_2$H$^+$ depletion toward the dust centre of TMC-1C.
Note that SMM 4 is probably not warm enough for CO to evaporate from the grain
mantles ($\sim20$ K, e.g., \cite{aikawa2008a}), so it is unlikely that CO, 
which is the main destroyer of N$_2$H$^+$ (through reaction
${\rm N_2H^+} + {\rm CO}\rightarrow {\rm HCO^+} + {\rm N_2}$),
would have led to disappearance of N$_2$H$^+$ from the gas phase. 

To study the chemical differentiation within the clump, we compare 
our previously determined NH$_3$ column densities with the present
N$_2$H$^+$ column densities. The integrated NH$_3$ $(1,1)$ and
$(2,2)$ intensity maps of the clump (see Appendix A in Harju et al. (1993))
show roughly the same morphology as the submm map. The NH$_3$ column densities
toward IRAS 05405-0117, Ori B9 E, and Ori B9 N are
$11.9\pm1.3\cdot10^{14}$, $7.2\pm1.3\cdot10^{14}$, and $9.7\pm4.6\cdot10^{14}$
cm$^{-2}$, respectively.
The corresponding NH$_3$/N$_2$H$^+$ column density ratios
are about $130\pm14$, $159\pm37$, and $236\pm87$. 
These values suggest that NH$_3$/N$_2$H$^+$ abundance ratio is higher towards
starless condensations than towards the IRAS source. Hotzel et al. (2004) found
a similar tendency in B217 and L1262: the NH$_3$/N$_2$H$^+$ abundance ratios
are at least twice as large in the dense starless parts of the cores
than in the regions closer to the YSO (see \cite{caselli2002b} for other
low-mass star-forming regions).
The same trend is also found in the high-mass star-forming region
IRAS 20293+3952 (\cite{palau2007}).
This is in accordance with chemistry models (\cite{aikawa2005}) and previous
observations (\cite{tafalla2004}) which suggest that NH$_3$ develops slightly
later than N$_2$H$^+$, and can resist depletion up to higher densities. 
It should be noted that models by Aikawa et al. (2005) reproduce the
observed enhancement of the NH$_3$/N$_2$H$^+$ ratio by adopting 
the branching ratio for the dissociative recombination of N$_2$H$^+$ as 
measured by Geppert et al. (2004; i.e., 
${\rm N_2H}^++{\rm e}\rightarrow{\rm NH}+{\rm N}$ accounts for 64\% of the 
total reaction). However, this branching ratio has been 
retreated by the same authors\footnote{Their recent laboratory experiment 
suggest that the above mentioned branching ratio is only 10\% (see 
\cite{aikawa2008b}).}, and thus it is not clear at the moment what
is actually causing the increase of NH$_3$/N$_2$H$^+$ abundance ratio.

\subsection{Core evolution: quasi-static vs. dynamic}

The degree of ionization in dense cores determines the importance of magnetic
fields in the core dynamics.
The ionization fractions in low-mass cores are found to be 
$10^{-8} < x({\rm e}) < 10^{-6}$ (\cite{caselli1998}; \cite{williams1998}).
The physical origin of the large variations in $x({\rm e})$ is not well
understood, though variations in $\zeta_{\rm H_2}$ or appropriate values
of metal depletion are assumed (\cite{padoan2004}). Padoan et al. (2004)
suggested that the observed variations in $x({\rm e})$ can be understood as the
combined effect of variations in core age, extinction, and density.

Fractional ionizations can be transformed into estimates of the
ambipolar diffusion (AD) timescale, $\tau_{\rm AD}$. We have used Eq.~(5)
of Walmsley et al. (2004) for this purpose. Assuming that H$^+$ is the
dominant ion (see Sect. 4.4), one obtains 
$\tau_{\mathrm{AD}}\sim8\cdot10^{13}x({\rm e}) \ \mathrm{yr}$.
If the dominant ion is HCO$^+$, 
$\tau_{\mathrm{AD}}\sim1.3\cdot10^{14}x({\rm e}) \ \mathrm{yr}$, i.e., 
$\sim60\%$ longer than in the former case.
Using the electron abundance $x({\rm e})=1\cdot10^{-7}$ we obtain that
$\tau_{\rm AD}\sim10^7$ yr. 
This timescale is roughly 70 and 100 times longer than the free-fall time 
($\tau_{\rm ff}\sim3.7\cdot10^7\left(n({\rm H_2}) [{\rm cm^{-3}}]\right)^{-1/2}$ yr) of Ori B9 E and N, respectively.
Since $\tau_{\rm AD}>\tau_{\rm ff}$, the cores may be supported against 
gravitational collapse by magnetic fields and ion-neutral coupling. 
The magnetic field that is needed to 
support the cores can be estimated using the relation between the critical
mass required for collapse and the magnetic flux 
(see Eq.~(2) in Mouschovias \& Spitzer (1976)). Using the masses and radii from
Table~\ref{table:parameters}, we obtain a critical magnetic field strength
of $\sim80$ $\mu$G for Ori B9 E/IRAS 05405-0117 and $\sim200$ $\mu$G for
Ori B9 N. These are rather high values compared to those that have been
observed (\cite{troland1996}; \cite{crutcher1999}; \cite{crutcher2000};
\cite{crutcher2004}; \cite{turner2006}; \cite{troland2008}).

According to the ``standard'' model of low-mass star formation,
$\tau_{\rm AD}/\tau_{\rm ff}\sim10$ (see, e.g., \cite{shu1987}; 
\cite{ciolek2001} and references therein). 
Since AD is generally a slow process, the core evolution
toward star formation occur quasi-staticly. 
The chemical abundances found in the present study 
($x({\rm N_2H^+})\sim10^{-10}$, $x({\rm NH_3})\sim10^{-7}$, see Sect. 5.8)
are consistent with chemical models for a dynamically young, but chemically
evolved (age $>10^5$ yr) source (\cite{bergin1997}; \cite{roberts2004};
\cite{aikawa2005}; \cite{shirley2005}; see also \cite{kirk2007}
and references therein). This supports the idea that the sources have
been static or slowly contracting for more than $10^5$ yr, and conforms with
the estimated AD timescales.

On the other hand, the equal numbers of prestellar and protostellar cores
suggest that the prestellar core lifetime should be similar 
to the lifetime of embedded protostars. Since the duration of the protostellar 
stage is $\sim{\rm few}\cdot10^5$ yr
(e.g., \cite{ward-thompson2007}; \cite{hatchell2007};
\cite{galvan-madrid2007}; \cite{enoch2008}),
the prestellar core evolution should be rather dynamic and last for only
a few free-fall times, as is the case in star formation driven by supersonic
turbulence (e.g., \cite{maclow2004}; \cite{ballesteros-paredes2007}).
This seems to contradict with the above results of 
AD timescales. However, in order to recognise the cores 
in the submm map, they are presumed to be in the high-density stage of
their evolution. Thus, the short statistical lifetime deduced above is still
consistent with the quasi-static evolution driven by AD,
if we are only observing the densest stages of a longer scale core evolution
(e.g., \cite{enoch2008}; \cite{crutcher2009}). Also, the dynamic phase in the
core evolution with $\tau_{\rm AD}$ being only a few free-fall times might be
appropriate for magnetically near-critical (the mass-to-magnetic flux ratio
being $\sim80\%$ of the critical value) or already slightly supercritical cores
when rapid collapse ensues (\cite{ciolek2001}; see also \cite{tassis2004}). 

\subsection{Linewidths and turbulence}

The N$_2$D$^+$ linewidths in Ori B9 E and N are significantly narrower
than the N$_2$H$^+$ linewidths (by factors of $\sim1.5$ and $\sim1.9$,
respectively). Crapsi et al. (2005a) found similar trend
in several low-mass starless cores (see their Table~4).
This is probably due to the fact that N$_2$D$^+$ traces the high density nuclei
of starless cores, where non-thermal turbulent motions are expected
to be insignificant (e.g., \cite{andre2007}; \cite{ward-thompson2007}).

The non-thermal component dominate the N$_2$H$^+$ linewidths in the observed
positions (the thermal linewidth of N$_2$H$^+$ is about 0.126 km s$^{-1}$ at
10 K, and thus $\Delta {\rm v}_{\rm NT}/\Delta {\rm v}_{\rm T}\sim2$).
However, the level of internal turbulence, as estimated from the ratio 
between the non-thermal velocity dispersion and the isothermal
speed of sound (e.g., \cite{kirk2007}), is not dynamically significant. 

Using Eq.~(7) of Williams et al (1998) and the derived values for the
molecular/metal ion-contribution constant $C_i$ and cosmic ray ionization rate
$\zeta_{\rm H_2}$ (see Table~\ref{table:ion}), we see that non-thermal
N$_2$H$^+$ line broadening in the observed positions can be
explained in part by magnetohydrodynamic (MHD) wave propagation.
The large wave coupling parameter in our sources ($W\gg1$), suggest that the 
coupling between the field and gas is strong and the waves are not suppressed.
The derived values of $W$ ($\sim30-60$, in the case of minimum turbulence)
are in agreement with $\tau_{\rm AD}/\tau_{\rm ff}$ ratios 
(\cite{williams1998}). 
Also, the estimated degrees of coupling between the magnetic field and gas
conforms with the susceptibility to fragmentation (\cite{bergin1999}).

Caselli \& Myers (1995) analysed ammonia cores in the Orion B GMC
and found an inverse relationship between core linewidth and distance to the
nearest stellar cluster. The nearest stellar cluster to Ori B9 is NGC 2024
at the projected distance of 5.2 pc (see Sect. 1.1.), so its role as driving
external turbulence to the region is probably not significant. 

\subsection{Formation of a small stellar group in Ori B9}

Internal turbulence or gravitational motions in the massive molecular cloud
core may promote fragmentation of the medium. This can easily generate 
sheets and filaments (e.g., \cite{caselli1995}; \cite{andre2008}). 
The collapse of these elongated clumps most probably results in the 
formation of a small stellar group or a binary system rather than a single
star (e.g., \cite{launhardt1996}). 
Only the densest parts of the filaments, the dense cores, are directly
involved in star formation.
It is unclear at the present time whether the collapse of an individual
prestellar core typically produces single stars or multiple protostellar
systems (see \cite{andre2008}).

The total mass of gas and dust of the clump associated with IRAS 05405-0117 
as derived from the dust continuum
emission is $\sim14$ M$_{\sun}$, and it has elongated structure with multiple
cores (local maxima in the filament are separated by more than one beam size,
see Fig.~\ref{figure:positions}).
The previous mass estimates by Harju et al. (1993) based on NH$_3$ were much
higher: $\sim50$ M$_{\sun}$ derived from $N({\rm NH_3})$ distribution, and
$\sim310$ M$_{\sun}$ derived from peak local density. 
The uncertainty in the abundance\footnote{Harju et al. assumed that 
$x({\rm NH_3})\sim3\cdot10^{-8}$. Using the H$_2$ column densities from the
dust continuum we derive the values of $x({\rm NH_3})\sim1-2\cdot10^{-7}$ in
our line observation positions.} and lower resolution used are certainly
affecting the estimation of the mass from NH$_3$.
However, the clump has enough mass to form a small stellar group. 

The kinetic temperature, velocity dispersion and the
fractional H$_2$D$^+$ abundance in the clump are similar to those
in the well-studied prestellar cores, e.g., L1544 and L183, where strong
emission of H$_2$D$^+$ line has been detected previously (see \cite{harju2006}
and references therein).
The masses, sizes, relatively high degree of deuteration and
the line parameters of the condensations indicate that they are low- to
intermediate-mass dense cores (cf. \cite{fontani2008}).
IRAS 05405-0117 and SMM 4 are likely to represent Class 0 protostellar cores
(see Sect. 5.1), whereas the subsidiary cores, e.g., Ori B9 N, are in an 
earlier, prestellar phase. 

\section{Summary and conclusions}

We mapped the Ori B9 cloud in the 870 $\mu$m dust continuum emission
with the APEX telescope. We also observed N$_2$H$^+(1-0)$ and N$_2$D$^+(2-1)$
spectral line emission towards selected positions in Ori B9 with the IRAM 30 m
telecope. These observations were used together with archival Spitzer/MIPS
data to derive the physical characteristics of the cores in Ori B9 and 
the degree of deuterium fractionation and ionization degree within the
IRAS 05405-0117 clump region. The main results of this work are:

\hspace{0.5cm}

1. The LABOCA field contains 12 compact submm sources. Four of them are 
previously known IRAS sources, and eight of them are new submm sources.
All the IRAS sources and two of the new submm sources are associated 
with the Spitzer 24 and 70 $\mu$m sources. The previously unknown
sources, SMM 3 and SMM 4, are promising Class 0 candidates based on their SEDs
between 24 and 870 $\mu$m. There is equal number of starless and
protostellar cores in the cloud. We suggest that the majority of our starless
cores are likely to be prestellar because of their high densities.

2. The total mass of the cloud as estimated from the 2MASS 
near-infrared extinction map is 1400 M$_{\sun}$.
The submm cores constitute about 3.6\% of
the total cloud mass. This percentage is in agreement with the observed low
values of star formation efficiency in nearby molecular clouds.

3. Mass distribution of the cores in Ori B9 and in Orion B North studied by 
Nutter \& Ward-Thompson (2007) very likely represent the subsamples of 
the same parent distribution. The CMF for the Orion B North 
is well-matched to the stellar IMF (\cite{goodwin2008}).
Also the core separations in these two regions are similar, indicating that 
the fragmentation length scale is similar. Since the fragmentation 
length scales are alike, and the CMFs have resemblance to the IMF, 
the origin of cores could be explained in terms of turbulent
fragmentation. The clustered mode of star formation in these two different
regions suggest that turbulence is driven on large scales.

4. On average, the starless cores are larger and more elongated than the 
protostellar cores in Ori B9. The observed mean angular sizes and axis
ratios suggest average density-power-law indices $p\sim1$ and $\sim1.5$ for 
starless and protostellar cores, respectively.

5. The fractional N$_2$H$^+$ and N$_2$D$^+$ abundances within the 
clump associated with IRAS 05405-0117 are $\sim4-11\cdot10^{-10}$ and 
$2-5\cdot10^{-11}$, respectively. The $N({\rm N_2D^+})/N({\rm N_2H^+})$ column
density ratio varies between 0.03-0.04. This is a typical degree of deuteration
in low-mass dense cores and conform with the earlier detection of H$_2$D$^+$. 
There is evidence for a N$_2$H$^+$ ``hole'' in the protostellar Class 0
candidate SMM 4. The envelope of SMM 4 probably represents an extreme case of
depletion where also N$_2$H$^+$ has disappeared from the gas phase.

6. The ionization fraction (electron abundance) in the
positions studied is estimated to be $x({\rm e})\sim10^{-7}$. There is 
uncertainty about the most abundant ionic species. The most likely candidates
are H$^+$ and HCO$^+$. The cosmic ray ionization rate in the
observed positions was found to be $\zeta_{\rm H_2}\sim1-2\cdot10^{-16}$
s$^{-1}$. 

7. There seems to be a discrepancy between the chemical age
derived near IRAS 05405-0117 and the statistical age deduced from the
numbers of starless and protostellar cores which suggest that
the duration of the prestellar phase of core evolution is
comparable to the free-fall time.
The statistical age estimate is, however, likely to be biased by the fact that 
the cores detected in this survey are rather dense ($n({\rm H_2})\gtrsim10^5$
cm$^{-3}$) and thus represent the most advanced stages.

\begin{acknowledgements}

We thank the referee, Paola Caselli, for her insightful comments and
suggestions that helped to improve the paper.
The authors are grateful to the sfaff of the IRAM 30 m telescope, for 
their hospitality and help during the observations. We also thank the staff
at the APEX telescope site. We are very grateful to Edouard Hugo, Oskar Asvany,
and Stephan Schlemmer for making available their rate coefficients of the 
reaction H$_3^+$ + H$_2$ with deuterated isotopologues. 
O. M. acknowledges Martin Hennemann for providing
the SED fitting tool originally written by
J\"urgen Steinacker, and the Research Foundation of
the University of Helsinki. The team acknowledges support from the Academy of
Finland through grant 117206.
This work is based in part on observations made
with the Spitzer Space Telescope, which is operated by the Jet Propulsion
Laboratory, California Institute of Technology under a contract with NASA.

\end{acknowledgements}

\end{document}